\documentclass[letterpaper, prd, aps, superscriptaddress, preprint, tightenlines]{revtex4-1}
\pdfoutput=1

\usepackage{amsmath,amssymb}
\usepackage{wasysym}
\usepackage{graphicx}
\usepackage{subcaption}
\usepackage[font=small]{caption}
\usepackage{multirow}
\usepackage{xcolor}
\usepackage{hyperref}
\hypersetup{
    colorlinks,
    linktoc=page,
    linkcolor={blue!70!black},
    citecolor={blue!70!black},
    urlcolor={blue!70!black}
}

\usepackage{bookmark}
\bookmarksetup{
  open,
  numbered
}

\newcommand{\mps}{M_\mathrm{PS}}
\newcommand{\mpi}{M_\pi}
\newcommand{\fps}{f_\mathrm{PS}}
\newcommand{\fpi}{f_\pi}
\newcommand{\mpcac}{m_\mathrm{PCAC}}
\newcommand{\mcrit}{m_\mathrm{crit}}
\newcommand{\csw}{c_\mathrm{sw}}

\newcommand{\mev}{\mathrm{MeV}}
\newcommand{\tr}{\mathrm{Tr}}
\newcommand{\reci}[1]{\frac{1}{#1}}

\newcommand{\cyp}{CaSToRC, The Cyprus Institute, 2121 Aglantzia, Nicosia, Cyprus}
\newcommand{\ucy}{Department of Physics, University of Cyprus, P.O. Box 20537, 1678 Nicosia, Cyprus}
\newcommand{\rmii}{Dip. di Fisica, Universit{\`a} and INFN di Roma Tor Vergata, 00133 Roma, Italy}
\newcommand{\fer}{Centro Fermi, Piazza del Viminale 1, 00184 Roma, Italy}
\newcommand{\nic}{NIC, DESY, Zeuthen, Platanenallee 6, 15738 Zeuthen, Germany}
\newcommand{\hub}{Institut f\"ur Physik, Humboldt-Universit\"at zu Berlin, Newtonstr. 15, 12489 Berlin, Germany}
\newcommand{\gre}{Theory Group, Lab. de Physique Subatomique et de Cosmologie, 38026 Grenoble, France}
\newcommand{\bon}{HISKP (Theory), Rheinische Friedrich-Wilhelms-Universit{\"a}t Bonn, Nu{\ss}allee 14-16, 53115 Bonn, Germany}
\newcommand{\ber}{Albert Einstein Center for Fundamental Physics, University of Bern, 3012 Bern, Switzerland}
\newcommand{\mnz}{Institut f{\"u}r Kernphysik, Johannes-Gutenberg-Universit{\"a}t Mainz, Johann-Joachim-Becher-Weg 45, 55128 Mainz, Germany}

\begin{document}

\title{\boldmath First Physics Results at the Physical Pion Mass from $N_f=2$ Wilson Twisted Mass Fermions at Maximal Twist}
\date{\today}

\author{A.~Abdel-Rehim}
\affiliation{\cyp}

\author{C.~Alexandrou}
\affiliation{\cyp}
\affiliation{\ucy}

\author{F.~Burger}
\affiliation{\nic}

\author{M.~Constantinou}
\affiliation{\cyp}
\affiliation{\ucy}

\author{P.~Dimopoulos}
\affiliation{\rmii}
\affiliation{\fer}

\author{R.~Frezzotti}
\affiliation{\rmii}

\author{K.~Hadjiyiannakou}
\affiliation{\cyp}

\author{C.~Helmes}
\affiliation{\bon}

\author{K.~Jansen}
\affiliation{\nic}

\author{C.~Jost}
\affiliation{\bon}

\author{C.~Kallidonis}
\affiliation{\cyp}

\author{B.~Knippschild}
\affiliation{\bon}

\author{B.~Kostrzewa}
\email[Corresponding author: ]{bartosz.kostrzewa@desy.de}
\altaffiliation[Current address: ]{\bon}
\affiliation{\nic}
\affiliation{\hub}

\author{G.~Koutsou}
\affiliation{\cyp}
\affiliation{\ucy}

\author{L.~Liu}
\affiliation{\bon}

\author{M.~Mangin-Brinet}
\affiliation{\gre}

\author{K.~Ottnad}
\altaffiliation[Current address: ]{\mnz}
\affiliation{\bon}

\author{M.~Petschlies}
\affiliation{\bon}

\author{G.~Pientka}
\affiliation{\hub}

\author{G.C.~Rossi}
\affiliation{\rmii}
\affiliation{\fer}

\author{C.~Urbach}
\affiliation{\bon}

\author{U.~Wenger}
\affiliation{\ber}

\author{M.~Werner}
\affiliation{\bon}

\collaboration{\textbf{ETM Collaboration}}
\noaffiliation

\begin{abstract}

We present physics results from simulations of QCD using $N_f=2$ dynamical Wilson twisted mass fermions at the physical value of the pion mass.
These simulations were enabled by the addition of the clover term to the twisted mass quark action.
We show evidence that compared to previous simulations without this term, the pion mass splitting due to isospin breaking is almost completely eliminated.
Using this new action, we compute the masses and decay constants of pseudoscalar mesons involving the dynamical up and down as well as valence strange and charm quarks at one value of the lattice spacing, $a\approx0.09\ \mathrm{fm}$.
Further, we determine renormalized quark masses as well as their scale-independent ratios, in excellent agreement with other lattice determinations in the continuum limit.
In the baryon sector, we show that the nucleon mass is compatible with its physical value and that the masses of the $\Delta$ baryons do not show any sign of isospin breaking.
Finally, we compute the electron, muon and tau lepton anomalous magnetic moments and show the results to be consistent with extrapolations of older ETMC data to the continuum and physical pion mass limits.
We mostly find remarkably good agreement with phenomenology, even though we cannot take the continuum and thermodynamic limits.\\
\vspace{0.1cm}

\noindent Preprint: DESY 15-121
\end{abstract}

\maketitle

\tableofcontents

\clearpage

\section{Introduction}\label{sec:introduction}

In the last decade the field of lattice QCD has seen significant progress in controlling systematic uncertainties. 
Advances in algorithms and lattice formulations have made it possible to study the continuum limit and the quark mass dependence of many phenomenologically interesting observables. 
Most recently, simulations with the physical value of the average up/down quark mass were performed making extrapolations to the physical pion mass superfluous, thereby eliminating the associated uncertainties.
An incomplete list of examples for observables related to the results presented in this paper can be found in Refs.~\cite{Durr:2010vn, Bazavov:2013vwa, PhysRevD.90.074509, Metivet:2014bga}.

The Wilson twisted mass formulation of lattice QCD (tmLQCD)~\cite{Frezzotti:2000nk} is one of a number of improved formulations with many advantages. 
Most importantly, tuned to maximal twist, leading lattice artifacts are of $\mathcal{O}(a^2)$ in physical observables~\cite{Frezzotti:2003ni}. 
However, twisted mass Wilson and standard Wilson fermions share a complicated phase structure~\cite{Farchioni:2004us, Farchioni:2004ma, Farchioni:2004fs, Farchioni:2005tu}: at finite values of the lattice spacing a remnant of the continuum chiral phase transition can render simulations with small values of the pion mass difficult. 
This phenomenon was predicted in Wilson chiral perturbation theory~\cite{Sharpe:1998xm, Munster:2003ba, Sharpe:2004ny, Scorzato:2004da} for Wilson type  fermions and found to occur in practice in Refs.~\cite{Baron:2009wt, Baron:2010bv}

In this paper we show in the $N_f=2$ case that these difficulties are overcome by adding the Sheikholeslami-Wohlert term~\cite{Sheikholeslami:1985ij} to the action.
This enables simulations at the physical pion mass with a value of the lattice spacing around $a=0.09\ \mathrm{fm}$ or even larger.
At maximal twist, all physical observables are automatically $\mathcal{O}(a)$ improved.
In contrast to Wilson clover fermions, a non-perturbative tuning of the Sheikholeslami-Wohlert coefficient $c_\mathrm{sw}$ is thus not needed.
Moreover, for any value of $c_\mathrm{sw}$, operator-specific improvement terms are not required in the maximally twisted theory.

With this new action, we have generated four ensembles with pion masses in the range from about $130\ \mathrm{MeV}$ to $500\ \mathrm{MeV}$ at one value of the lattice spacing.
On these gauge configurations, we compute pseudoscalar meson masses and decay constants as well as their ratios, the nucleon and $\Delta$ baryon masses, the anomalous magnetic moments of the electron, muon and tau leptons and a number of gluonic scales.
The meson mass ratios are used to give estimates of renormalised quark masses as well as their scale-independent ratios.

Our results show that the aforementioned quantities can be extracted with good statistical precision. 
We believe, therefore, that simulations at the physical pion mass together with automatic $\mathcal{O}(a)$ improvement with this action will provide high precision results for phenomenologically interesting observables such as quark masses, weak matrix elements and the hadronic contribution to the anomalous magnetic moment of the muon. 
In fact, a comparison of the results obtained with the new action at one value of the lattice spacing to continuum extrapolated results from previous $N_f=2$ simulations reveals only small deviations, with the exception of $D$ meson related quantities for which it is well known that a continuum extrapolation is essential. 
The ensembles discussed in this paper are also the basis for results on meson and nucleon structure~\cite{Abdel-Rehim:2015owa}, some of which are the first obtained directly at the physical pion mass.

The paper is organised as follows: in section~\ref{sec:action} we detail the lattice action and the observables investigated. 
Section~\ref{sec:results} is devoted to the results, followed by a summary in section~\ref{sec:summary}. 
Details about the simulation parameters and the analysis procedure are given in the Appendices~\ref{app:simdetails} and~\ref{app:meson_details}. 
 
\section{Lattice Action and Observables}
\label{sec:action}

For the discretised gauge action we use the so-called Iwasaki gauge action~\cite{Iwasaki:1985we} as used for the previous $N_f=2+1+1$ ETMC simulations. 
Compared to previous simulations performed by ETMC the fermion action has been modified by adding the so-called clover term~\cite{Sheikholeslami:1985ij} to read:
\begin{equation}
  \label{eq:sf}
  S_\ell^\mathrm{tm} = \sum_x \bar{\chi}_\ell\left[ D_W(U) + m_0 + i \mu_\ell \gamma^5 \tau^3 +
    \frac{i}{4} \csw \sigma^{\mu\nu}
    \mathcal{F}^{\mu\nu}(U)  \right] \chi_\ell(x)\,,
\end{equation}
where $D_W$ is the massless Wilson Dirac operator, $m_0$ the bare Wilson mass parameter, $\mu_\ell$ is the bare twisted mass parameter and $c_{\mathrm{sw}}$ is the so-called Sheikoleslami-Wohlert improvement coefficient \cite{Sheikholeslami:1985ij}. 
$\tau^3$ is the third Pauli matrix acting in flavour space and $\bar{\chi}_\ell,\chi_\ell$ are the fermionic fields in the twisted basis $\chi_\ell=(u,d)^t$. 
We remark that we will work in the twisted basis throughout this paper unless stated otherwise.

The bare Wilson mass $m_0$ is tuned to its critical value $\mcrit$ by requiring the PCAC quark mass
\begin{equation}
  \label{eq:mpcac}
  \mpcac =
  \frac{\sum_\mathbf{x}\langle\partial_0A^a_0(\mathbf{x},t)P^a(0)\rangle}  {2\sum_\mathbf{x}\langle P^a(\mathbf{x},t)P^a(0)\rangle} \,
  ,\qquad\quad a=1,2\, 
\end{equation}
to vanish at every value of the bare twisted mass $\mu_\ell$ independently. 
Here $A_\mu^a$ and $P^a$ are the axial vector current and the pseudo scalar density in the twisted basis, respectively, 
\[
A_\mu^a(x) = \bar\chi_\ell(x)\gamma_\mu\gamma_5\frac{\tau^a}{2}\chi_\ell(x)\,
,\qquad\qquad P^a(x) = \bar\chi_\ell(x)\gamma_5\frac{\tau^a}{2}\chi_\ell(x)\, .
\]
In this situation -- called maximal twist -- physical observables are free of $\mathcal{O}(a)$ lattice artefacts without the need of any improvement coefficients~\cite{Frezzotti:2003ni}. 

The clover term is usually introduced in order to obtain on-shell $\mathcal{O}(a)$ improvement of lattice QCD with Wilson fermions~\cite{Luscher:1996ug} by tuning $\csw$ non-per\-tur\-ba\-ti\-ve\-ly using a suitable condition in the massless theory. 
Since in our case $\mathcal{O}(a)$ improvement is already guaranteed by Wilson twisted mass at maximal twist, we can use the clover term to modify artefacts of $\mathcal{O}(a^2)$ and, possibly, reduce them. 

In particular, it was shown in the quenched approximation~\cite{Becirevic:2006ii, Dimopoulos:2009es} that combining Wilson twisted mass fermions at maximal twist and the clover term reduces cut-off effects related to isospin symmetry breaking in twisted mass lattice QCD. 
Following Ref.~\cite{Becirevic:2006ii}, we have set $\csw$ to its non-perturbative value $\csw=1.57551$ using Pad{\'e} fits to data in Ref.~\cite{Aoki:2005et}. 
We stress again that it is not necessary to use the non-perturbative value and in principle $\csw$ can be tuned by requiring minimal mass splitting between the charged and the neutral pion.
In addition, it must be noted that all the symmetries which ensure automatic $\mathcal{O}(a)$-improvement at maximal twist persist when a clover term is present.

\begin{table}[t!]
  \small
  \centering
  \begin{tabular*}{1.\textwidth}{@{\extracolsep{\fill}}lccccccccc}
    \hline\hline
    ensemble & $\beta$ & $\csw$ & $\kappa_c$ & $a\mu_\ell$ &
    $L/a$ & $N_\tau$ & $N_\mathrm{conf}$ & $\tau$ & $\tau_\mathrm{int}(P)$ \\
    \hline\hline
    \textit{cA2.09.48}   & $2.10$ & $1.57551$ & $0.13729$ & $0.0009$ & $48$ &
    $6900$ & $2950$ & $1$ & $15(6)$ \\
    \textit{cA2.30.24}   & $2.10$ & $1.57551$ & $0.13730$ & $0.0030$ & $24$ &
    $3400$ & $1300$ & $1$ & $3.2(8)$ \\
    \textit{cA2.60.24}   & $2.10$ & $1.57551$ & $0.13730$ & $0.0060$ & $24$ &
    $9000$ & $4000$ & $1$ & $3.8(6)$ \\
    \textit{cA2.60.32}   & $2.10$ & $1.57551$ & $0.13730$ & $0.0060$ & $32$ &
    $11840$ & $5350$ & $1$ & $2.9(5)$ \\
    \hline\hline
  \end{tabular*}
  \caption{The ensembles used in this investigation, all of which have temporal extent $T=2L$ with $L/a$ the spatial lattice extend.
    In addition we give the total number of trajectories $N_\tau$, the number of thermalised configurations $N_\mathrm{conf}$ and the HMC trajectory length $\tau$ and the integrated autocorrelation time of the plaquette $\tau_\mathrm{int}(P)$. }
  \label{tab:ens}
\end{table}

The gauge configurations have been generated using the Hybrid Monte-Carlo (HMC) algorithm with mass preconditioning and multiple time scales~\cite{Urbach:2005ji}. The corresponding code is publicly available in the tmLQCD software suite~\cite{Jansen:2009xp, Abdel-Rehim:2013wba, Deuzeman:2013xaa}.  
The ensemble details are summarised in Table~\ref{tab:ens} including the number of configurations, the number of trajectories and the HMC trajectory length. Configurations have been saved every second trajectory after a suitable number of equilibration trajectories. 

For ensemble $\mathit{cA2.09.48}$, the bare twisted mass has been tuned such that the ratio $\mpi/\fpi$ takes its physical value.
A detailed listing of all simulation parameters for all ensembles and a discussion of molecular dynamics histories is given in Appendix~\ref{app:simdetails}.

Quantities with strange and charm quark content are probed on our $N_f=2$ flavour ensembles by adding valence strange and charm quarks in the so-called Osterwalder-Seiler (OS) discretisation~\cite{Frezzotti:2004wz}. 
The corresponding fermionic action for a doublet of OS flavours $f\in \{s,c\}$ with bare twisted masses $a\mu_{s,c}$ reads
\begin{equation}
  \label{eq:Sos}
  S^\mathrm{OS}_f = \bar{\chi}_f \left[ D_W(U) + m_0 + i
    \mu_f \gamma^5 \tau^3 +
    \frac{i}{4} \csw \sigma^{\mu\nu}
    \mathcal{F}^{\mu\nu}(U)\right]\chi_f\,. 
\end{equation}
Formally, the action in Eq.~\ref{eq:Sos} is accompanied by a corresponding ghost action to exactly cancel their sea contribution. 
For more details we refer to Ref.~\cite{Frezzotti:2004wz}.
When $m_0$ is set equal to the value of $\mcrit$ of the unitary action, $\mathcal{O}(a)$ improvement stays valid for arbitrary values of $\csw$.

\subsection{Lattice Scales from Gluonic Observables}
\label{subsec:gluonic scales}

We begin by discussing our determinations of various lattice scales from gluonic observables and hence not specific to twisted mass fermions. 
We consider two types of scales, namely one related to the static quark-antiquark potential $r_0$, and the ones related to the action density renormalised through the gradient flow~\cite{Luscher:2010iy}.

The gradient flow $B_\mu(t,x)$ of gauge fields is defined in the continuum by the flow equation
\begin{align}
\label{eq:GF equation continuum 1}
\dot B_\mu &= D_\nu G_{\nu\mu}, \quad \left. B_\mu \right|_{t=0} =
A_\mu\, ,\\
G_{\mu\nu} &= \partial_\mu B_\nu - \partial_\nu B_\mu + [B_\mu,B_\nu],
\quad D_\mu = \partial_\mu + [B_\mu, \cdot]\,,
\label{eq:GF equation continuum 2}
\end{align}
where $A_\mu$ is the fundamental gauge field, $G_{\mu\nu}$ the field strength tensor and $D_\mu$ the covariant derivative.
At finite lattice spacing Eqs.~\ref{eq:GF equation continuum 1} and \ref{eq:GF equation continuum 2} become 
\begin{equation}
 \frac{d}{dt} V_t(x,\mu) = -g_0^2 \cdot {\partial_{x,\mu} S_G(V_t)} \cdot V_t(x,\mu)\,,
\label{eq:GF equation lattice}
\end{equation}
where $V_t(x,\mu)$ is the flow of the original gauge field $U(x,\mu)$ at flow time $t$, $S_G$ is an arbitrary lattice discretisation of the gauge action and $\partial_{x,\mu}$ denotes the su$(3)$-valued differential operator with respect to $V_t$. 
For our calculations we use the standard Wilson gauge action. 
One virtue of the gradient flow is that observables evaluated on gauge fields at flow times $t>0$ are renormalised~\cite{Luscher:2011bx}. 
One can, therefore, define lattice scales by keeping a suitable renormalised gluonic observable, e.g.~the action density $E$~\cite{Luscher:2010iy}, at constant flow time $t_0$ fixed in physical units, through the condition
\begin{equation}
t_0^2 \langle E(t_0)\rangle = E_0
\label{eq:GF t0 definition}
\end{equation}
and determine the lattice scale from the dimensionless flow time in lattice units, $\hat t_0 = a^2 t_0$. 
For convenience we will also sometimes use $\hat s_0 = \sqrt{\hat t_0}$. For our calculation we use both the standard Wilson plaquette 
\begin{equation}
 E_\text{pl}(t) = 2 \sum_{p \in P_x} \text{Re} \, \text{tr} \{1 - V_t(p)\}\,,
\end{equation}
and a symmetrised clover-like discretisation for the action density $E_\text{sym}$~\cite{Luscher:2010iy}. The difference between the results from the two definitions can be used to estimate the size of the effects stemming from the discretisation of the action density.

An alternative scale $w_0$ has been introduced in Ref.~\cite{Borsanyi:2012zs} and is defined through a suitable derivative of the action density,
\begin{equation}
\label{eq:GF w0 definition}
W(t) = t \cdot \partial_t \left(t^2 \langle E(t)\rangle\right) \, ,
\end{equation}
and the condition
\begin{equation}
W(t=w_0^2) = W_0 \, .
\end{equation}

In addition to the lattice scales from $t_0, s_0$ and $w_0$ we also consider the scale from the dimensionful combination $t_0/w_0$. 
The combination has been found to have a very weak dependence on the quark mass~\cite{Deuzeman:2012jw}. Because the scales from the gradient flow of the action density are strongly correlated, they should not be regarded as independent. 
In particular, correlations need to be taken into account in the combination $t_0/w_0$. 
Moreover, since the action density at $t \sim t_0 \sim w_0^2$ usually suffers from large autocorrelation~\cite{Deuzeman:2012jw}, the calculation of the statistical error needs special care.

An independent scale can be calculated from the static quark-antiquark potential. 
In this approach, a scale is defined through the force $F(r)$ between a static quark and antiquark separated by the distance $r$~\cite{Sommer:1993ce}. 
The condition
\begin{equation}
\label{eq:r0 definition}
r_0^2 F(r_0) = 1.65
\end{equation}
fixes the scale $\hat r_0 = r_0/a$. 
The static force can be determined from the static quark-antiquark potential $V(r)$ through the calculation of Wilson loops. 
More specifically, the potential at distance $r$ is extracted from the asymptotic time dependence of the $r \times t$-sized Wilson loops $W(r,t)$,
\begin{equation}
\label{eq:V(r) Wilson loops}
\lim_{t\rightarrow \infty} \langle W(r,t) \rangle \propto
e^{-V(r)\cdot t} \, ,
\end{equation}
and the force is then determined through the derivative of a suitable parametrisation of the potential as a function of $r$ which we choose as
\begin{equation}
V(r) = V_0 + \frac{\alpha}{r} + \sigma r \, .
\label{eq:V parametrization}
\end{equation}
In order to optimise the overlap of the Wilson loop with the ground state of the potential, we employ five different levels of spatial APE-smearing and extract the ground state energy from the corresponding correlation matrix by solving the corresponding generalised eigenvalue problem~\cite{Niedermayer:2000yx}. 
Finally, we also make use of the noise reduction proposed in Ref.~\cite{DellaMorte:2003mn}. 
Further details on the calculation of the Wilson loops and the analysis procedure can be found in Ref.~\cite{Boucaud:2008xu} and Ref.~\cite{Niedermayer:2000yx}.
 
\subsection{Pseudo-scalar Meson Masses and Decay Constants}\label{subsec:pseudoscalar}

We continue with a discussion of the masses and decay constants of pseudo-scalar mesons such as pions, kaons, D- and D$_s$-mesons. 
We define the pseudo-scalar interpolating operator for flavours $f,f'\in \{\ell,s,c\}$ as 
\begin{equation}
  P^\pm_{f,f'}(t)\ =\ \sum_\mathbf{x}\bar{\chi}_f(\mathbf{x},t)\, i\gamma_5\,
  \tau^\pm\, \chi_{f'}(\mathbf{x},t)\,,\qquad \tau^\pm =
  \frac{\tau^1\pm i\tau^2}{2}
\end{equation}
and the pseudo-scalar correlation function
\begin{equation}
  \label{eq:Cps}
  C_\mathrm{PS}^{f,f'}(t)\ =\ \langle\, P^\pm_{f,f'}(t)\ P^\pm_{f,f'}(0)^\dagger\, \rangle\,.
\end{equation}
This choice ensures that flavours $f$ and $f'$ always come with opposite values of their corresponding twisted mass parameters. 
In the light sector, this choice projects to the charged pion states.
In the kaon and D-meson case in principle also the combinations with equal signs of light and $s$ or $c$ quarks are possible, because they lead to the same meson mass values and amplitudes in the continuum limit. 
However, one can show that in the case of opposite signs leading cut-off effects in the squared pseudo-scalar meson masses are of $\mathcal{O}(m_f a^2)$ with $m_f$ the relevant quark mass~\cite{Frezzotti:2005gi, Sharpe:2004ny}.

The spectral decomposition of $C_\mathrm{PS}^{f,f'}(t)$ allows one to extract the pseudo-scalar meson mass $\mps^{f,f'}$ from
\[
\lim_{t\to\infty} C_\mathrm{PS}^{f,f'}(t) = \frac{|\langle 0| P^\pm_{f,f'}
| \mathrm{PS}\rangle|^2}{2\mps^{f,f'}}\ (e^{-\mps^{f,f'}t} + e^{-\mps^{f,f'}(T-t)})\,,
\]
where $|\mathrm{PS}\rangle$ is the ground state in this channel. 
$\mps^{\ell,\ell}, \mps^{\ell,s}$ and $\mps^{\ell,c}$ correspond to the charged pion, the kaon and the D-meson masses, respectively. 
Let us also define the effective mass
\begin{equation}
  \label{eq:effmass}
  M_\mathrm{eff}(t)=-\log(C(t)/C(t-1))
\end{equation}
for general correlation functions $C(t)$, which can also be utilised to determine hadron masses.
The matrix element $\langle 0| P^\pm_{f,f'}| \mathrm{PS}\rangle$ is at maximal twist directly related to the pseudo-scalar decay constant via
\begin{equation}
\label{eq:fps}
\fps^{f,f'}\ =\ (\mu_f+\mu_{f'})\frac{\langle 0| P^\pm_{f,f'}| \mathrm{PS}\rangle}{(\mps^{f,f'})^2}\,, 
\end{equation}
which follows from the PCVC relation in Wilson twisted mass lattice QCD at maximal twist~\cite{Frezzotti:2003ni}.
The lattice dispersion relation for mesons can be taken into account by exchanging ${(\mps^{f,f'})^2}$ in Eq.~\ref{eq:fps} for $\mps^{f,f'} \sinh(\mps^{f,f'})$.
In the following, the former will be referred to as ``Continuum Definition'' (CD) and the latter as ``Lattice Definition'' (LD).

Due to flavour symmetry breaking in Wilson twisted mass lattice QCD, charged and neutral pions differ in their mass values by $\mathcal{O}(a^2)$ artifacts. 
Reducing this mass splitting -- and, therefore, allowing simulations at the physical point -- was one of the main design goals of the action specified in Eq.~\ref{eq:sf}.
The mass of the neutral pion can be determined from the interpolating operator in the twisted basis
\begin{equation}
  \label{eq:P0}
  P^0(t) = \sum_\mathbf{x} \bar\chi_\ell (\mathbf{x}, t)\, \mathbf{1}_\mathrm{F}\, \chi_\ell(\mathbf{x},t) \,,
\end{equation}
where we denote with $\mathbf{1}_\mathrm{F}$ the unit matrix in flavour space.
The corresponding correlation function $C_\mathrm{PS}^0(t)$ has connected and disconnected contributions and is, therefore, noisy.
In the following we denote the charged pion mass as $\mpi$, the full neutral one as $M_{\pi^0}$ and the one determined from only the connected part of $C_\mathrm{PS}^0(t)$ as $M_{\pi^{(0,c)}}$.
The techniques used to extract the full neutral pion mass with sufficient statistical precision are detailed in Appendix~\ref{app:neutralpion}.

In order to extract the ground state masses and matrix elements more reliably, we include also fuzzed~\cite{Lacock:1994qx} interpolators in our analysis. 
From local and fuzzed interpolators we build a $2\times2$ matrix and solve the corresponding GEVP~\cite{Michael:1982gb, Luscher:1990ck, Blossier:2009kd} or use a constrained matrix fit~\cite{Boucaud:2008xu}. 
For further details we refer to Ref.~\cite{Boucaud:2008xu}.

\subsection{Nucleon and Delta masses}\label{subsec:nucleon}

The mass of the nucleon  is extracted from two-point correlators using the standard interpolating fields, which are given in the physical quark field basis for the proton by
\begin{equation} 
  \label{eq:Proton}
  J_p = \epsilon_{abc}\bigl( u^T_a C\gamma_5 d_b\bigr)u_c\,,
\end{equation}
where $C=\gamma_4\gamma_2$ denotes the charge conjugation matrix and spinor indices are suppressed.
For the $\Delta^{++}$ and $\Delta^{+}$, we use the interpolating fields 
\begin{eqnarray}
  \label{eq:Delta}
  J^\mu_{\Delta^{++}} &=& \epsilon_{abc}\bigl( u^T_a C\gamma^\mu u_b\bigr)u_c, \\
  J^\mu_{\Delta^{+}} &=& \frac{1}{\sqrt{3}}\epsilon_{abc}\biggl[
  2\bigl( u^T_a C\gamma^\mu d_b\bigr)u_c +\bigl( u^T_a C\gamma^\mu u_b\bigr)d_c
  \biggl]\,.
\end{eqnarray}

In order to improve the overlap with  the ground state we employ gauge invariant smearing that has been demonstrated to effectively suppress excited state contributions.  
Gaussian smearing~\cite{Gusken:1989qx, Alexandrou:1992ti} is applied to each  quark field,  $q({\bf x},t)$  yielding a  smeared quark field, $q^{\rm smear}({\bf x},t) = \sum_{\bf y} F({\bf x},{\bf y};U(t)) q({\bf y},t)$.
The gauge invariant smearing function is given by
\begin{equation}
  F({\bf x},{\bf y};U(t)) = (1+\alpha H)^ n({\bf x},{\bf y};U(t)),
\end{equation}
constructed from the hopping matrix understood as a matrix in coordinate, color and spin space,
\begin{equation}
  H({\bf x},{\bf y};U(t))= \sum_{i=1}^3 \biggl( U_i({\bf x},t)\delta_{{\bf x,y}-a\hat \imath} +  U_i^\dagger({\bf x}-a\hat \imath,t)\delta_{{\bf x,y}+a\hat \imath}\biggr).
\end{equation}
The parameters $\alpha$ and $n$ are varied so that the root mean square (r.m.s) radius obtained using the proton interpolating field is of the order of 0.5~fm.
The values $\alpha=4$ and $n=50$ are seen to produce an early plateau for the effective mass in Eq.~\ref{eq:effmass}, where the appropriate correlation function is the zero-momentum two-point correlator of the proton
\begin{equation}
  \label{eq:C_P}
  C_p(t) = \frac 1 2 {\rm Tr}(1 \pm \gamma_4) \sum_{\bf x}
  \langle J_p( {\bf x}, t) \bar J_p(0, 0)\rangle\,.
\end{equation}
In addition, we apply APE smearing~\cite{Albanese:1987ds} to the spatial links that enter the hopping matrix in the smearing function, setting $\alpha_{\rm APE}=0.5$ and $n_{\rm APE}=50$.
APE smearing is useful to reduce the gauge noise in the correlation functions.

The interpolating field for the $\Delta$ has also overlap to spin-1/2 states. 
This overlap can be removed with the incorporation of a spin-3/2 projector in the definitions of the interpolating fields
\begin{equation} 
  J_{X_{3/2}}^\mu = P^{\mu\nu}_{3/2} J^\nu_{X} \, .
\end{equation}
For non-zero momentum, $P^{\mu\nu}_{3/2}$ is defined as~\cite{Benmerrouche:1989uc}
\begin{equation} \label{eq:proj32}
  P^{\mu\nu}_{3/2} = \delta^{\mu\nu} - \reci{3}\gamma^\mu \gamma^\nu - \reci{3p^2}\left(\not{p}\gamma^\mu p^\nu + p^\mu \gamma^\nu \not{p}  \right) \, .
\end{equation}
The spin-1/2 component $J_{X_{1/2}}^\mu$ can be obtained by acting with the spin-1/2 projector $P^{\mu\nu}_{1/2}=\delta^{\mu\nu} - P^{\mu\nu}_{3/2}$ on $J^\mu_X$. 
Components with Lorentz indices $\mu,\nu=0$ will not contribute. 
Since we are interested in the mass we take $\mathbf{p}=0$ in which case the last term of Eq.~\ref{eq:proj32} will contain $\delta_{0\mu}$ and vanish. 
When the spin-3/2 and spin-1/2 projectors are applied to the interpolating field operators, the resulting two-point correlators for the spin-3/2 and 1/2 baryons acquire the form
\begin{eqnarray} 
  \label{eq:correlators_32_12}
  C_{\frac{3}{2}} (t) &=& \frac{1}{3}\tr [C(t)] + \frac{1}{6} \sum_{i\ne j}^3 \gamma_i \gamma_j C_{ij}(t)\;, \nonumber\\
  C_{\frac{1}{2}} (t) &=& \frac{1}{3}\tr [C(t)] - \frac{1}{3} \sum_{i\ne j}^3 \gamma_i \gamma_j C_{ij}(t)\;,
\end{eqnarray}
where $\tr[C] = \sum_i C_{ii}$~\cite{Alexandrou:2014sha}. 
When no projector is taken into account, the resulting two-point correlator would be equal to $\frac{1}{3}\tr[C]$.
Although for the $\Delta$ the contribution from the spin-1/2 component is suppressed~\cite{Alexandrou:2008tn}, we nevertheless include the spin-3/2 projector.

\subsection{Anomalous Magnetic Moments}\label{subsec:g-2}

The leading-order hadronic contribution to the lepton anomalous magnetic moments in Euclidean space-time is given by~\cite{Blum:2002ii}
\begin{equation}
  a_{\mathrm{l}}^{\mathrm{hvp}} = 
  \alpha^2 \int_0^{\infty} \frac{d Q^2 }{Q^2} 
  w\left( \frac{Q^2}{m_{\mathrm{l}}^2}\right) \Pi_{\mathrm{R}}(Q^2) \; ,
  \label{eq:amudef}
\end{equation}
where $\alpha$ is the fine structure constant, $w$ is a weight function, $Q^2$ is the squared Euclidean momentum and $m_{\mathrm{l}}$ is the corresponding lepton mass. 
$\Pi_{\mathrm{R}}(Q^2)$ is the renormalised hadronic vacuum polarisation function, $\Pi_{\mathrm{R}}(Q^2)= \Pi(Q^2)- \Pi(0) $, obtained from the vacuum polarisation tensor 
\begin{equation}
  \label{eq:vptensor}
  \Pi_{\mu \nu}(Q)= \int d^4 x \,e^{iQ\cdot(x-y)} \langle J_{\mu}^\mathrm{em}(x) J_{\nu}^\mathrm{em}(y)\rangle = (Q_{\mu} Q_{\nu} - Q^2
  \delta_{\mu
    \nu}) \Pi(Q^2)\,,
\end{equation}
with the electromagnetic vector current $J_\mu^\mathrm{em}(x)$. 
For the lattice computation of the quark-connected diagrams contributing to $a_{\mathrm{l}}^{\mathrm{hvp}}$ we employ the conserved point-split vector current for a single flavour $q$, 
\begin{equation}
 J_{\mu}(x)  =  \frac{1}{2} \left( \overline{q}(x+\hat{\mu})(1+\gamma_{\mu}) U_{\mu}^{\dagger}(x) q(x)  -
\overline{q}(x)(1-\gamma_{\mu}) U_{\mu}(x) q(x+\hat{\mu}) \right)\,.
\label{eq:conscurrent}
\end{equation}

The hadronic vacuum polarisation function defined as in Ref.~\cite{Burger:2014ada} is fitted by dividing the momentum squared range between $0$ and $100\,{\rm GeV}^2$ in a low-momentum region $ 0 \le Q^2 \le Q_\mathrm{match}^2$ and a high-momentum one $Q_\mathrm{match}^2 < Q^2 \le Q_\mathrm{max}^2=100\ \mathrm{GeV}^2$ according to
\begin{equation}
  \label{eq:pilowandhigh}
  \Pi(Q^2) = (1- \Theta(Q^2-Q^2_{\rm match}))\Pi_{\mathrm{low}}(Q^2) + \Theta(Q^2-Q^2_{\rm match}) \Pi_{\mathrm{high}}(Q^2) \,,
\end{equation}
where the low-momentum fit function is chosen to be
\begin{equation}
  \label{eq:pilow}
  \Pi_{\mathrm{low}}(Q^2) = \sum_{i=1}^M \frac{f^2_i}{M^2_i + Q^2} + \sum_{j=0}^{N-1} a_j (Q^2)^{j} \,,
\end{equation}
and the form of the high-momentum part is inspired by perturbation theory 
\begin{equation}
  \label{eq:pihigh}
  \Pi_{\mathrm{high}}(Q^2) = \log(Q^2) \sum_{k=0}^{B-1} b_k (Q^2)^{k}  + \sum_{l=0}^{C-1} c_l (Q^2)^{l} \,. 
\end{equation}
This defines our so-called MNBC fit function, e.g. M1N2B4C1 means $M=1$, $N=2$, $B=4$, and $C=1$ in Eqs.~\ref{eq:pilow} and \ref{eq:pihigh} above.
$M_i$ and $f_i$ represent the energy levels and decay constants in the vector channel, respectively, which are determined from the corresponding two point functions, see Ref.~\cite{Burger:2014ada}.
$a_i$, $b_i$ and $c_i$ are free parameters to be fitted to the data.
As the value of $Q^2_{\rm match}$ in the Heaviside functions in Eq.~\ref{eq:pilowandhigh} we have chosen $2\ \mathrm{GeV}^2$. 
Varying the value of $Q^2_{\rm match}$ between $1\,{\rm GeV}^2$ and $3\,{\rm GeV}^2$ does not lead to observable differences as long as the transition between the low- and the high-momentum part of the fit is smooth. 

Since the momentum, where the weight function appearing in the definition of $a_{\mathrm{l}}^{\rm hvp}$ in Eq.~\ref{eq:amudef} attains its maximum, is proportional to the squared lepton mass and the lepton masses vary over four orders of magnitude, the different lepton anomalous magnetic moments are sensitive to very different momentum regions. 
Thus, the different lepton moments provide a very valuable cross-check of the interpolation method we used.

\section{Results}
\label{sec:results}
In this section we present results for a number of observables determined on the ensembles used in this study.
We remark that we currently have only one value of the lattice spacing and one volume at the physical point available.
Therefore, we cannot control the associated systematic errors, which will be addressed in the future.

\subsection{Lattice Scales from Gluonic Observables}
\label{subsec:gluonic results}

In Table~\ref{tab:results gluonic scales} we compile our results for the various gluonic scales discussed in section~\ref{subsec:gluonic scales}. 
The integrated autocorrelation times $\tau_\text{int}^{E(t_0)}$ refer to the action density evaluated at flow time $t_0$ and our estimates should be understood as lower bounds for the true values. 

\begin{table}[thb]
  \begin{tabular*}{\textwidth}{l @{\extracolsep{\fill}} lllll}
    \hline\hline \\[-2.0ex]
    ensemble & $\tau_\text{int}^{E(t_0)}$  & $t_0/a^2$ & 
    $(t_0/w_0)/a$  & $w_0/a$ &$r_0/a$ \\[1.0ex]
    \hline \hline \\[-2.0ex]
    \textit{cA2.09.48} & 37(16)& 2.8037(23)&  1.50964(50) &  1.8572(14)& 5.317(48)    \\
    \textit{cA2.30.24} & 15(6) & 2.8022(85)&  1.5134(11)    &  1.8517(55)& 5.322(114)   \\
    \textit{cA2.60.24} & 12(5) & 2.7404(73)&  1.5105(15)   &  1.8142(41)& 5.162(53)    \\
    \textit{cA2.60.32} & 23(5) & 2.7367(28)&  1.5111(03)    &  1.8110(17)&  5.191(63)   \\
    \hline\hline
  \end{tabular*}
  \caption{Results for the gluonic scales $r_0/a$, $t_0/a^2$ and $w_0/a$ as well as the integrated autocorrelation time of the action density at flow time $t_0$.}
  \label{tab:results gluonic scales}
\end{table}

The results for the scales from the gradient flow are based on the symmetrised action density $E_\text{sym}$. 
All errors take into account the autocorrelation times either by blocking or by including explicitly $\tau_\text{int}^{E(t_0)}$ which is given in units of HMC trajectories. 
The error on the combination $t_0/w_0$ takes the strong correlation between $t_0$ and $w_0$ into account by a correlated bootstrap analysis. The correlation reflects itself in a very small relative statistical error of 0.3\permil\ which is less than half the relative error on $\sqrt{t_0/a^2}$ or $w_0/a$. What makes the scale $t_0/w_0$ even more compelling is the circumstance that its quark mass dependence is very weak and in fact negligible within statistical errors. This has already been observed for our $N_f=2$ twisted mass ensembles using the twisted mass Dirac operator at maximal twist without the clover term \cite{Deuzeman:2012jw} and it would be interesting to investigate this independence in view of the $\chi$PT expressions provided in Ref.~\cite{Bar:2013ora}. For the other two scales $t_0$ and $w_0$ we observe a shift of about 2.5\% from the result at $a\mu=0.006$ compared to the one at the physical point. 
From the last two rows in Table~\ref{tab:results gluonic scales} we can further draw the conclusion that finite volume effects in the gluonic lattice scales from the gradient flow are also negligible for the volumes considered here. Finally, an estimate of the intrinsic lattice artefacts in the scales can be obtained by comparing the results above with the corresponding ones based on the action density calculated from the plaquette. The difference between the two definitions is about 10\% for $t_0/a^2$ and $(t_0/w_0)/a$ while it is only about 1.5\% for $w_0/a$, but of course these numbers do not say much about the true lattice artefacts in the scales listed in Table \ref{tab:results gluonic scales}.

The total error on $r_0/a$ results from the statistical and systematic errors added in quadrature. 
The estimates of the systematic errors in $r_0/a$ due to neglected excited state contributions, interpolating $F(r)$ to $r_0$ and lattice artefacts are obtained as follows. The excited state contributions are estimated from the shift in $r_0/a$ when repeating the whole analysis with all temporal fit ranges shifted by one unit. The interpolation error is estimated from the variation of $r_0/a$ under a change of the interpolation range. The lattice artefacts are estimated from the shift in $r_0/a$ when the analysis is repeated using Iwasaki improved distances for the potential instead of the naive ones. Below, the separate statistical, excited state, interpolation and lattice artefact errors can be read off in that order
\begin{align*}
r_0/a(\text{\textit{cA2.09.48}}) &= 5.317(08)(01)(25)(40)\,, \\ 
r_0/a(\text{\textit{cA2.30.24}}) &= 5.322(52)(89)(38)(32)\,,\\
r_0/a(\text{\textit{cA2.60.24}}) &= 5.162(24)(31)(08)(34)\,,\\
r_0/a(\text{\textit{cA2.60.32}}) &= 5.191(13)(22)(38)(43)\,.
\end{align*}
From the comparison of the values for the two volumes available for {\it cA2.60}, we conclude that the finite volume effects in $r_0/a$ are well within the statistical error. However, we do observe a quark mass dependence for $r_0$ yielding a difference of about 2.5\% between the result at $a\mu=0.006$ and the one at the physical point, which is of the same size as the shift in $t_0$ and $w_0$.
 
\subsection{Simulation Stability and Twisted Mass Isospin Breaking} 
\label{subsec:mass-splitting}

Before moving on to present our results for masses and decay constants, we discuss the stability of the simulations with the new action and the effects of the explicit isospin symmetry breaking with twisted mass fermions.
The biggest (and almost only, see also Ref.~\cite{Dimopoulos:2009qv}) effect of this isospin breaking in past simulations has always been observed in the charged to neutral pion mass splitting.
This splitting is responsible for the lower bound, $\mu_{\mathrm{crit}}$, on the bare light quark mass value which can be simulated at given value of the lattice spacing~\cite{Sharpe:1998xm, Munster:2003ba, Sharpe:2004ny, Scorzato:2004da}.
In fact, meta-stabilities were observed in simulations with $\mu<\mu_\mathrm{crit}$~\cite{Farchioni:2004us, Farchioni:2004fs, Farchioni:2005tu}.

\begin{figure}
 \centering
 \begin{subfigure}{0.46\linewidth}
 \includegraphics[width=\linewidth,page=1]{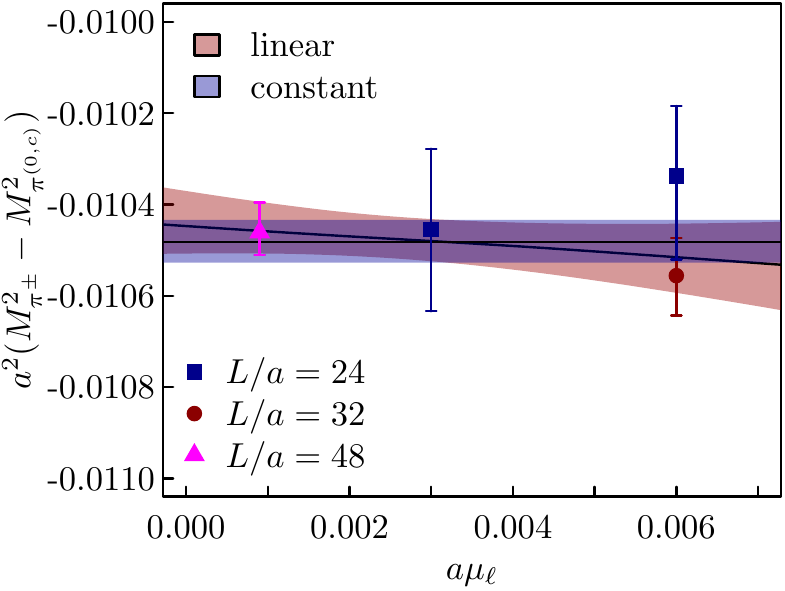}
 \caption{}\label{fig:mass-splitting-chiral:mpi_mpi0c}
 \end{subfigure}
 \hfill
 \begin{subfigure}{0.46\linewidth}
 \includegraphics[width=\linewidth,page=1]{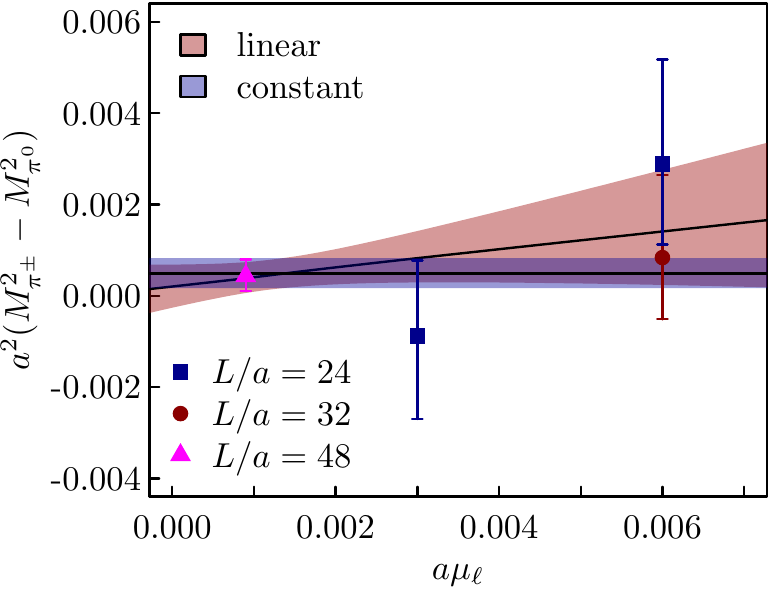}
 \caption{}\label{fig:mass-splitting-chiral:mpi_mpi0}
 \end{subfigure}
 \caption{Charged and neutral pion mass splittings \textbf{(a)}: $a^2 (M^2_{\pi^\pm} - M^2_{\pi^{(0,c)}}) $ and \textbf{(b)}: $ a^2 ( M^2_{\pi^\pm} - M^2_{\pi^0}) $ as a function of the bare light quark mass $a \mu_\ell$.}
 \label{fig:mass-splitting-chiral}
\end{figure}

In Table~\ref{tab:pions_afK_afD_afDs_aMDs}, we list the charged, connected neutral and full neutral pion masses in lattice units.
The determination of the full neutral pion mass is somewhat subtle due to the presence of a vacuum expectation value as well as disconnected diagrams, as discussed in Appendix~\ref{app:neutralpion}.
As shown in Figure~\ref{fig:mass-splitting-chiral:mpi_mpi0}, linear and constant extrapolations in the light quark mass of the splitting between the squared full neutral and charged pion masses to the chiral limit are found to be
\begin{align}
  a^2 ( M_{\pi^\pm}^2 - M_{\pi^0}^2 )^{\mathrm{lin}}_{a\mu_\ell \to 0} & = 0.00018(^{+32}_{-28}) \\
  a^2 ( M_{\pi^\pm}^2 - M_{\pi^0}^2 )^{\mathrm{cst}}_{a\mu_\ell \to 0} & = 0.00026(^{+21}_{-22}) \, ,
\end{align}
clearly compatible with zero.
We can thus conclude that the low energy constant $c_2$ parametrizing the $\mathcal{O}(a^2)$ lattice artefact responsible for the pion mass splitting is zero within our uncertainties.
At the physical charged pion mass, taking the extremal values of the errors as an upper bound, the difference between the charged and neutral pion masses is no larger than about $13$ MeV.
 
In addition, the splitting between the charged and the connected neutral pion -- as shown in panel~\ref{fig:mass-splitting-chiral:mpi_mpi0c} -- is seen to be about a factor of three smaller than measured on ensembles~\cite{Herdoiza:2013sla} without the clover term.
We take these as strong indications that the new action has significantly reduced the isospin breaking compared to previous simulations, even at the rather coarse lattice spacing employed here. 
This is in line with quenched studies~\cite{Becirevic:2006ii} which were the main motivation for proceeding with twisted mass clover fermions at maximal twist.

\begin{table}
  \footnotesize
  \centering
  \begin{tabular*}{1.\textwidth}{@{\extracolsep{\fill}}lllll}
    \hline\hline \\[-2.0ex]
    observable                              & \textit{cA2.09.48}           & \textit{cA2.30.24}          & \textit{cA2.60.24}          & \textit{cA2.60.32}          \\[0.6ex]
    \hline\hline \\[-2.0ex]
    $aM_{\pi^\pm}$                          & $0.06196(09)(^{+12}_{-05})$  & $0.1147(7)(^{+4}_{-7})$     & $0.15941(38)(^{+15}_{-21})$ & $0.15769(26)(^{+15}_{-14})$ \\[0.6ex]
    $aM_{\pi^{(0,c)}}$                      & $0.1191(05)(^{+07}_{-10})$   & $0.1541(13)(^{+05}_{-05})$  & $0.18981(61)(^{+21}_{-25})$ & $0.18840(44)(^{+46}_{-29})$ \\[0.6ex]
    $aM_{\pi^0}$                            & $0.0593(27)(^{+16}_{-11})$   & $0.1163(55)(^{+51}_{-12})$  & $0.1489(50)(^{+41}_{-53})$  & $0.1554(34)(^{+20}_{-57})$ \\[0.6ex]
    $af_{\pi^\pm}^\mathrm{(CD)}$            & $0.06042(11)(^{+07}_{-03})$  & $0.06104(43)(^{+15}_{-14})$ & $0.06946(22)(^{+03}_{-05})$ & $0.07043(19)(^{+06}_{-05})$ \\[0.6ex]
    $af_{\pi^\pm}^\mathrm{(LD)}$            & $0.06038(11)(^{+07}_{-03})$  & $0.06090(43)(^{+16}_{-14})$ & $0.06917(22)(^{+03}_{-05})$ & $0.07013(19)(^{+06}_{-05})$ \\[0.6ex]
    $M_{\pi^\pm}/f_{\pi^\pm}^\mathrm{(CD)}$ & $1.0254(31)(^{+26}_{-12})$   & $1.879(22)(^{+08}_{-17})$   & $2.30(11)(^{+02}_{-03})$    & $2.2395(76)(^{+39}_{-24})$  \\[0.6ex]
    $M_{\pi^\pm}/f_{\pi^\pm}^\mathrm{(LD)}$ & $1.0260(31)(^{+26}_{-12})$   & $1.884(22)(^{+09}_{-17})$   & $2.31(11)(^{+02}_{-03})$    & $2.2489(77)(^{+38}_{-23})$  \\[1.0ex]
    $af_K^\mathrm{(CD)}$                    & $0.07235(9)(^{+2}_{-2})$     & $0.07265(31)(^{+06}_{-06})$ & $0.07774(19)(^{+07}_{-07})$ & $0.07816(16)(^{+09}_{-07})$ \\[0.6ex]
    $af_K^\mathrm{(LD})$                    & $0.07173(9)(^{+2}_{-2})$     & $0.07197(32)(^{+06}_{-05})$ & $0.07692(19)(^{+07}_{-06})$ & $0.07734(16)(^{+09}_{-07})$ \\[0.6ex]
    $af_D^\mathrm{(CD)}$                    & $0.1022(9)(^{+3}_{-7})$      & $0.1087(14)(^{+09}_{-13})$  & $0.1127(7)(^{+5}_{-7})$     & $0.1110(10)(^{+05}_{-06})$  \\[0.6ex]
    $af_D^\mathrm{(LD)}$                    & $0.0906(8)(^{+2}_{-6})$      & $0.0960(12)(^{+07}_{-11})$  & $0.0994(6)(^{+4}_{-6})$     & $0.0980(8)(^{+4}_{-5})$     \\[0.6ex]
    $af_{D_s}^\mathrm{(CD)}$                & $0.1207(2)(^{+1}_{-1})$      & $0.1220(7)(^{+1}_{-1})$     & $0.1237(5)(^{+1}_{-2})$     & $0.1219(5)(^{+1}_{-1})$     \\[0.6ex]
    $af_{D_s}^\mathrm{(LD)}$                & $0.1058(2)(^{+1}_{-1})$      & $0.1068(5)(^{+1}_{-1})$     & $0.1082(4)(^{+1}_{-1})$     & $0.1067(5)(^{+1}_{-1})$     \\[0.6ex] 
    $aM_{D_s}$                              & $0.9022(27)(^{+06}_{-07})$   & $0.905(3)(^{+1}_{-1})$      & $0.9062(27)(^{+02}_{-02})$  & $0.9034(26)(^{+01}_{-01})$  \\[1.0ex]
    $N_\mathrm{meas} (N^{\pi_0}_\mathrm{meas}) $ & $1457 (615)$            & $728 (352)$                 & $1351 (424)$                & $670 (337)$ \\[0.6ex]
    \hline\hline
  \end{tabular*}
  \caption{Charged, neutral connected and full neutral pion masses as well as charged pion decay constants in lattice units and their ratios determined on the ensembles in this analysis. In addition, values in lattice units for a number of mesonic quantities. The first error is statistical and the second is an estimate of the systematic error due to the choice of fit range. $N_\mathrm{meas} (N^{\pi_0}_\mathrm{meas})$ represents the number of configurations used for meson quantities (the full neutral pion mass, respectively) on the corresponding ensemble.}
  \label{tab:pions_afK_afD_afDs_aMDs}
\end{table}

\begin{figure}
  \centering
  \begin{subfigure}{0.49\linewidth}
    \includegraphics[width=\linewidth]{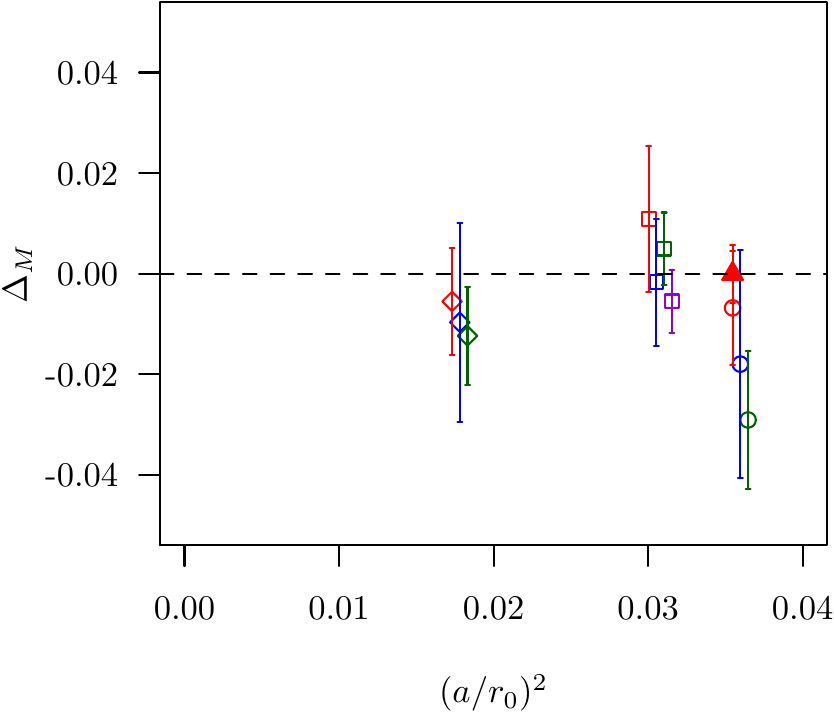}
    \caption{}\label{fig:mass-splitting-delta}
  \end{subfigure}
  \begin{subfigure}{0.49\linewidth}
    \includegraphics[width=\linewidth]{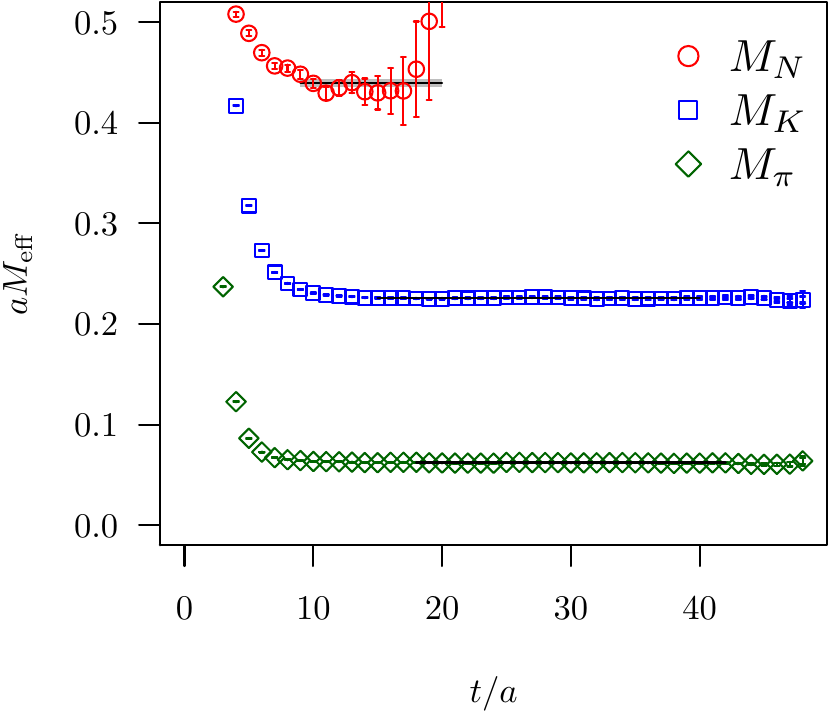}
    \caption{}\label{fig:effective-masses}
  \end{subfigure}
  \caption{\textbf{(a)} Relative mass differences $\Delta_M$ for the $\Delta$ baryons as a function of $(a/r_0)^2$ for $N_f=2+1+1$ ensembles (open circles for $\beta=1.90$, squares for $\beta=1.95$ and diamonds for $\beta=2.10$) as well as for the ensemble at the physical value of the pion mass (filled red triangle). 
    Red symbols represent the lightest pion mass, then blue, green and violet for increasing pion mass values for each lattice spacing. 
    The data points have been slightly displaced horizontally for legibility.
    \textbf{(b)} Effective masses as a function of $t/a$ of the nucleon, kaon and pion for the physical pion mass ensemble \textit{cA2.09.48.}}
  \label{fig:delta_massdiff}
\end{figure}

Let us now discuss the baryon sector.
There is still an exact lattice symmetry, namely parity combined with an interchange of $u$- with a $d$-quark, which means for example that the proton and the neutron are degenerate as are the $\Delta^{++}$ and the $\Delta^{-}$ as well as the $\Delta^+$ and $\Delta^0$. 
However, a mass splitting could be seen between the $\Delta^{++}$ and the $\Delta^+$. 
Thus, we average the mass of the $\Delta^{++}$ and $\Delta^-$ as well as that of the $\Delta^+$ and $\Delta^0$ and take the difference between the two averages as a measure of the magnitude of the isospin breaking.
We show in the left panel of Figure~\ref{fig:delta_massdiff} the dimensionless ratio
\[
\Delta_M = \frac{M_{\Delta^{++,-}}-M_{\Delta^{+,0}}}{M_{\Delta^{++,-}}}
\]
versus $(a/r_0)^2$ for the old $N_f=2+1+1$ ensembles as well as for the new ensemble \textit{cA2.09.48} at the physical pion mass. 
As can be seen, the splitting is consistent with zero, indicating that isospin breaking effects are small for the $\Delta$ baryons.  

\subsection{Hadron Masses and Decay Constants} 
\label{subsec:meson_results}

In this section we give results for pseudo-scalar meson masses, their decay constants and the nucleon mass.
The analysis procedure for the mesons is described in detail in Appendix~\ref{app:meson_details}, with particular emphasis on the estimate of the asymmetric systematic errors quoted in Tables~\ref{tab:mass_ratios} and~\ref{tab:ps_ratios}.
For the nucleon mass the error is at present still too large for a sensible analysis of systematic uncertainties.

As an example for the mass determinations, we show in the right panel of Figure~\ref{fig:delta_massdiff} the effective masses for the nucleon, the pion and the kaon, the latter with $a\mu_s=0.0245$. 
In this figure, for the nucleon the effective masses are computed from a smeared-smeared correlation function, while for the pion and kaon the local-local correlation functions were used.

In the case of the nucleon we clearly observe the expected exponential error growth.
This is why we do not include effective masses for $t/a>21$ in the analysis.
The fit result, error and range are indicated by the solid line and shaded region.
For the pion and kaon, a plateau is visible up to $t=T/2$, as expected. 
For the two pseudo-scalar particles the indicated fit range is only an example, because we perform a weighted average over many fit ranges, as explained in Appendix~\ref{app:fitrange}.
Errors on $\mpi$ and $M_K$ are too small to be visible on this scale, but details can be found in  Figure~\ref{fig:m_pi_K.effectivemass}, page~\pageref{fig:m_pi_K.effectivemass}.

\begin{figure}[!t]
  \centering
  \begin{subfigure}{0.49\linewidth}
    \includegraphics[width=\linewidth]{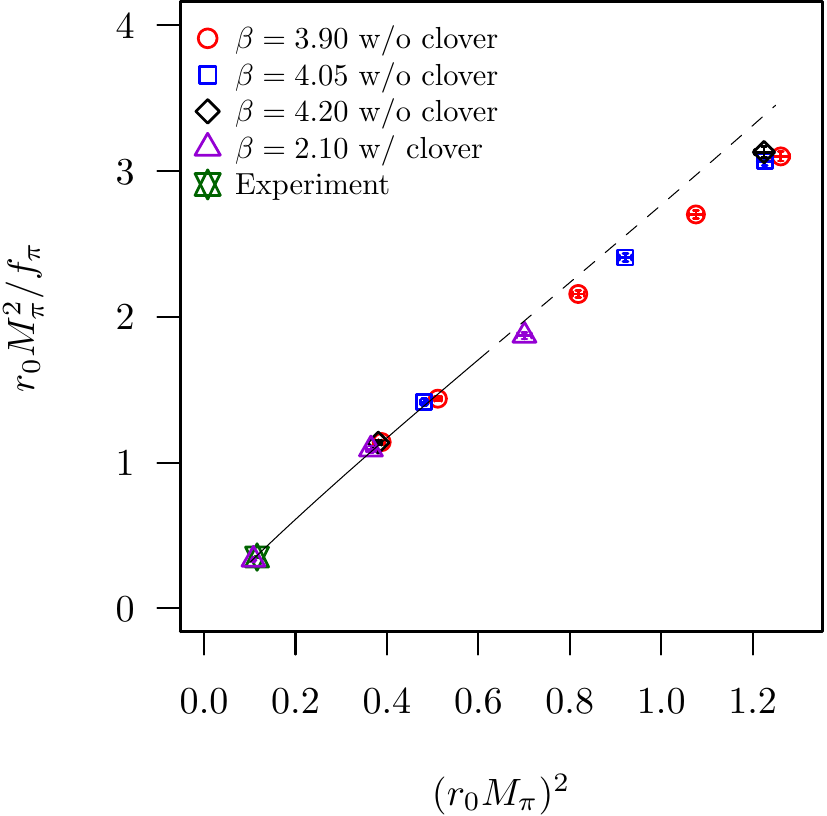}
    \caption{}
  \end{subfigure}
  \begin{subfigure}{0.49\linewidth}
    \includegraphics[width=\linewidth]{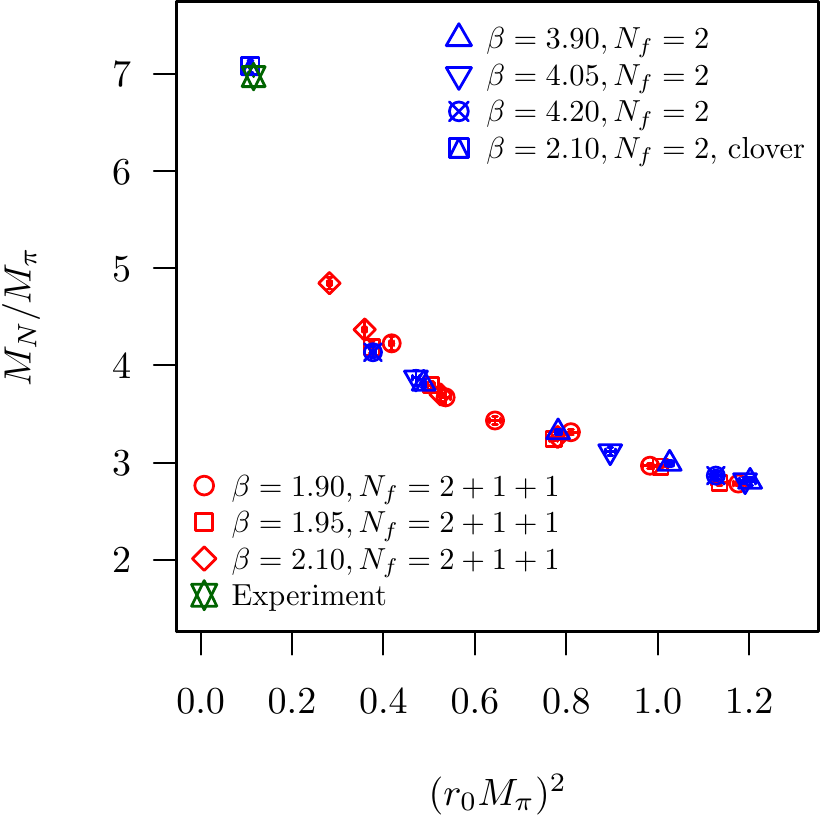}
    \caption{}
  \end{subfigure}
  \caption{\textbf{(a)} $r_0\mpi^2/\fpi$ as a function of $(r_0\mpi)^2$ comparing
    $N_f=2$ results w/o clover term~\protect{\cite{Baron:2009wt}} with
    the new results presented in this paper. The line is a
    NLO $\chi$PT fit to the data as explained in the text.
    \textbf{(b)} Ratio of the nucleon mass to the pion mass as a function of
    the pion mass squared in units of $r_0$. 
    We show data for $N_f=2$ w/o clover term, $N_f=2+1+1$ and the new physical point result.}
  \label{fig:mN}
\end{figure}

In the left panel of Figure~\ref{fig:mN} we show, as a function of $(r_0\mpi)^2$, the ratio $r_0 \mpi^2/\fpi$ in which some of the lattice artefacts seem to cancel.
In this plot we compare the new $N_f=2$ results presented in this paper with the $N_f=2$ results from ETMC simulations without clover term~\cite{Baron:2009wt}.
We show the bare data with only Gasser-Leutwyler finite size corrections~\cite{Gasser:1986vb} applied, together with the experimental value. 
In addition we show a fit of the NLO $\chi$PT expression~\cite{Weinberg:1978kz, Gasser:1983yg, Gasser:1985gg}
\[
\frac{\mpi^2}{\fpi}\ =\ \frac{\mpi^2}{f_0} \left(1 + 2\frac{\mpi^2}{(4\pi f_0)^2}\log\frac{\mpi^2}{\Lambda_4^2} \right)
\]
as a function of $\mpi^2$ to the data (excluding the experimental one) in units of $r_0$, neglecting lattice artefacts.
Restricting the fit range to $\mpi < 300\ \mathrm{MeV}$ (indicated by the solid line), we obtain $f_0 = 0.122(4)\ \mathrm{GeV}$ and $\bar\ell_4 \equiv \log M_{\pi^{\pm}}^2/\Lambda_4^2 = 3.3(4)$.
The $p$-value of this fit is $0.49$, and the inclusion of a chiral log is clearly favoured over a linear fit.
If one fits $\mpi/\fpi$ instead, the results do not change, only the $p$-value gets significantly worse, indicating residual finite volume and lattice artefacts.
These fit results are completely in line with the results of Ref.~\cite{Durr:2013goa}.

We remark that due to the smaller pion mass splitting with clover term in the action we expect also the finite size corrections to be smaller than for the ensembles without clover term.
Moreover, at the physical point finite size corrections are expected to be small, because they are proportional to $\mpi^2$.
Hence, corrections discussed in Refs.~\cite{Colangelo:2010cu, Bar:2010jk} are likely to give only tiny contributions.

The mass ratio of the nucleon to the pion for our $N_f=2$ and $N_f=2+1+1$ ensembles is shown in the right panel of Figure~\ref{fig:mN} as a function of the pion mass squared in units of $r_0$. 
The nucleon mass for the physical point has been measured on $96$ independent configurations with $16$ sources per configuration.
The masses for the $N_f=2$ ensembles without clover term have been taken from Ref.~\cite{Alexandrou:2009qu}, the ones for the $N_f=2+1+1$ ensembles from Ref.~\cite{Alexandrou:2014sha}.
The values for $r_0/a$ where taken from Ref.~\cite{Carrasco:2014cwa} for $N_f=2+1+1$ and from Ref.~\cite{Blossier:2010cr} for $N_f=2$ w/o clover term. 
As can be seen, the lattice results follow a universal curve indicating that cut-off effects are small on this ratio. 
Moreover, differences between $N_f=2$ and $N_f=2+1+1$ are smaller than the statistical uncertainties.

For quantities involving strange and charm quarks the valence quark mass needs to be tuned.
This tuning was performed by matching the phenomenological values of the pseudo-scalar meson mass ratios $M_K/M_\pi$ and $M_D/M_\pi$ through linear interpolations of the lattice data, resulting in the bare quark masses and their ratios given in Table~\ref{tab:mass_ratios}.
The details of this procedure are discussed in Appendix~\ref{app:mass_tuning}.

\begin{table}[h]
  \centering
  \begin{tabular}{ccc}
    \hline \hline \\[-2.0ex]
    $ a\mu_l$                  & $a\mu_s$                     & $a\mu_c$ \\[0.8ex]
    \hline \\[-1.8ex]
    0.0009                     & $0.02485(7)(^{+4}_{-3})$     & $0.3075(15)(^{+14}_{-14})$ \\[0.8ex]
    \hline \hline \\[-2.0ex]
    $\mu_s/\mu_l$              & $\mu_c/\mu_l$                & $\mu_c/\mu_s$ \\[0.8ex]
    \hline \\[-1.8ex]
    $27.61(8)(^{+4}_{-4})$     & $342.1(1.8)(^{+1.6}_{-1.6})$ & $12.39(8)(^{+6}_{-9})$ \\[1.0ex]
    \hline \hline 
  \end{tabular}
  \caption{Bare quark masses in lattice units and their ratios as determined by matching $M_K/M_\pi$ and $M_D/M_\pi$ to their phenomenological values. For $\mu_c/\mu_s$, the asymmetric error is computed by considering the maximum spread of the asymmetric errors on the dividend and divisor while for $\mu_s/\mu_l$ and $\mu_c/\mu_l$ it comes from $M_K/M_\pi$ and $M_D/M_\pi$ directly.}
  \label{tab:mass_ratios}
\end{table}

Ratios of meson masses and decay constants resulting from this analysis are given in Table~\ref{tab:ps_ratios}.
It is clear from $M_\pi/f_\pi$ that the ensemble \textit{cA2.09.48} is at the physical pion mass within errors.
For the other quantities agreement with phenomenological determinations and continuum limit lattice averages is quite good.
As an estimate of the residual $\mathcal{O}(a^2)$ artefacts, one can compare the difference between the two definitions (see Eq.~\ref{eq:fps} and below) of the decay constant in quantities involving $f_D$ and $f_{D_s}$.
It seems that these effects should be no larger than about 15\%, indicating that a well-behaved continuum limit should certainly be achievable.
Finally, one can compare to ETMC determinations from Ref.~\cite{Blossier:2009bx} for $N_f=2$ twisted mass fermions in the infinite volume and continuum limit which gave $f_K/f_\pi = 1.210(18)$ and $f_{Ds}/f_D = 1.24(3)$, in excellent agreement with the present analysis.

\begin{table}
\begin{tabular*}{1.\textwidth}{@{\extracolsep{\fill}}cllll}
\hline \hline \\[-2.0ex]
observable         & CD & LD & PDG~\cite{Agashe:2014kda} & FLAG~\cite{2013:FlagReview} \\[0.8ex]
\hline \hline \\[-2.0ex]
$M_\pi/f_\pi$      & $1.0254(31)(^{+26}_{-12})^\dagger$        & $1.0262(30)(^{+33}_{-18})^\dagger$        & $1.0337(28)^\star$        & $1.035(11)^\star$ \\[0.6ex]
$M_K/f_K$          & $3.1404(55)(^{+13}_{-11})^\dagger$        & $3.1675(56)(^{+13}_{-11})^\dagger$        & $3.164(14)^\star$         & $3.162(18)^\star$ \\[0.6ex]
$M_{D}/f_{D}$      & $8.395(64)(^{+41}_{-16})^\dagger$         & $9.466(71)(^{+41}_{-17})^\dagger$         & $9.11(22)$                & --                \\[0.6ex]     
$M_{D_s}/f_{D_s}$  & $7.474(21)(^{+03}_{-03})$                 & $8.531(28)(^{+04}_{-03})$                 & $7.64(14)$                & --                \\[0.6ex]
$M_{D_s}/M_\pi$    & $14.564(54)(^{+03}_{-03})^\dagger$        & --                                        & $14.603(33)^\star$        & --                \\[0.6ex]
$f_K/f_\pi$        & $1.1976(21)(^{+06}_{-07})$                & $1.1881(21)(^{+06}_{-07})$                & $1.1979(57)$              & $1.200(15)$       \\[0.6ex] 
$f_D/f_\pi$        & $1.694(14)(^{+04}_{-10})$                 & $1.503(12)(^{+04}_{-07})$                 & $1.569(38)$               & $1.61(3)$         \\[0.6ex]
$f_{D_s}/f_\pi$    & $1.998(6)(^{+1}_{-1})$                    & $1.751(5)(^{+1}_{-1})$                    & $1.975(35)$               & $1.91(3)$         \\[0.6ex]
$f_D/f_K$          & $1.413(12)(^{+02}_{-03})$                 & $1.264(10)(^{+02}_{-02})$                 & $1.309(33)$               & $1.34(2)$         \\[0.6ex]
$f_{D_s}/f_D$      & $1.206(23)(^{+04}_{-04})$                 & $1.190(22)(^{+04}_{-04})$                 & $1.258(38)$               & $1.19(2)$         \\[0.6ex]
\hline\hline
\end{tabular*}
\caption{Ratios of pseudo-scalar meson observables calculated on the gauge ensemble \textit{cA2.09.48} at the physical pion mass interpolated to the strange and charm valence quark masses from the matching procedure described in Appendix~\ref{app:mass_tuning}. 
We give results with CD and LD definitions (see Eq.~\ref{eq:fps}) of the decay constants separately, where applicable.
All starred reference ratios involving $M_\pi$ or $M_K$ use the isospin symmetric values of these quantities taken from Ref.~\cite{2013:FlagReview}. Daggered quantities are not independent and given for reference only. ``FLAG'' refers to $N_f=2+1$ averages.} 
 \label{tab:ps_ratios}
\end{table}
 
\subsection{Estimate of the Lattice Spacing}
\label{subsec:lattice_spacing}

In this section we provide estimates of the lattice spacing in physical units.
In order to simplify the discussion we concentrate on the results at the physical point. 
From the discussion in section~\ref{subsec:gluonic results} it is clear that the finite volume corrections in the gluonic scales are negligible, as well as the effects from possibly being slightly off the physical point.
The lattice values  we use  for $\sqrt{t_0/a^2}, (t_0/w_0)/a, w_0/a$ and $r_0/a$ are the ones given in the first row of Table~\ref{tab:results gluonic scales}.

First we determine the lattice spacing from the gluonic scales $\sqrt{t_0/a^2}, w_0/a$ and $r_0/a$. 
For the latter we refer to the summary of values given in Ref.~\cite{Sommer:2014mea} for $N_f=2$ QCD. 
There are three determinations by the CLS collaboration~\cite{Lottini:2013rfa,Capitani:2011fg} which yield $r_0^\text{CLS} = 0.4877(77)$ fm from a weighted average assuming 100\% correlation between the determinations. The correlations are taken into account by following the procedure by Schmelling~\cite{Schmelling:1994pz}.
Combined with the values from the ALPHA collaboration \cite{Fritzsch:2012wq} and the QCDSF collaboration~\cite{Bali:2012qs} we obtain the weighted average $r_0 = 0.4945(57)$ fm. 
In addition there are also three determinations from our earlier $N_f=2$ simulations using the tree-level Symanzik improved gauge action and twisted mass fermions without the clover term \cite{Blossier:2009bx,Baron:2009wt,Alexandrou:2009qu}. 
A weighted average assuming 100\% correlation yields $r_0^\text{ETMC} = 0.443(20)$ fm which exposes a sizable tension between the ETMC determinations and the ones by other collaborations. 
Nevertheless, we can also average the results from all collaborations and obtain the weighted average $r_0 = 0.4907(86)$ fm where the error is stretched by $\sqrt{\chi^2/\text{dof}} = 1.6$ in order to account for the incompatibility of the results. 
This is the value we use for our determination of the lattice spacing from $r_0$. 
For the physical values of $\sqrt{t_0}, w_0$ we refer to Ref.~\cite{Bruno:2013gha} which so far provides the only $N_f=2$ determinations. 
The values are $\sqrt{t_0} = 0.1535(12)$ fm and $w_0 = 0.1757(13)$ fm.
 
In Table \ref{tab:lattice_spacings_gluonic} we collect the physical values for the gluonic scales as discussed above together with the resulting lattice spacings calculated from our determinations at the physical point.
\begin{table}
 \centering
 \begin{tabular*}{.6\textwidth}{@{\extracolsep{\fill}}rrcccccc}
   \hline \hline
   & [fm] &  $a$ [fm] \\
   \hline
   $\sqrt{t_0}$ & 0.1535(12) & 0.0917(07) \\
   $w_0$        & 0.1757(13) & 0.0946(07) \\
   $r_0$        & 0.4907(86) & 0.0923(18) \\
   \hline
   avg.         &            & 0.0931(08) \\    
   \hline \hline
 \end{tabular*}
 \caption{Lattice spacing in physical units determined from gluonic quantities at the physical point using the input from Ref.~\protect{\cite{Bruno:2013gha}} given in the second row. The last line contains the weighted average of the three determinations assuming 100\% correlation.}
 \label{tab:lattice_spacings_gluonic}
\end{table}
In the last line we provide the weighted average of the lattice spacings together with the statistical error, assuming 100\% correlation between our gluonic lattice data for $\sqrt{t_0}/a$, $w_0/a$ and $r_0/a$.  
The weighted average  yields $a = 0.0931(08)$ fm with a $\chi^2/\text{dof}=4.7$. 
We assign the large value of $\chi^2/\text{dof}$ to lattice artefacts, since for the gluonic quantities this is the only systematic error which we do not control in our simulation at the physical point.  
In order to account for this uncertainty, we quote a systematic error covering the spread of the determinations. In this way we obtain
\begin{equation}
  a_\mathrm{gluonic} = 0.0931(08)(15) \, \text{fm} \, .
  \label{eq:gluonic lattice spacing}
\end{equation}

In addition to the gluonic scales in Table~\ref{tab:results gluonic scales} we can make use of hadronic quantities, namely the nucleon mass $M_N$, the pion mass $\mpi$ and the pion and kaon decay constants $f_\pi$ and $f_K$, respectively. 
The lattice values $a f_\pi=a f_{\pi^\pm}^\text{(CD)}, a \mpi = a M_{\pi^\pm}$ and $a f_K=a f_K^\text{(CD)}$ can be found in Table~\ref{tab:pions_afK_afD_afDs_aMDs} on page~\pageref{tab:pions_afK_afD_afDs_aMDs}. 
For the nucleon mass we use the value $a M_N = 0.440(4)$ \cite{Abdel-Rehim:2015owa}.
The physical values we use are taken from the 2014 edition of the PDG Review of Particle Physics~\cite{Agashe:2014kda}. 
Fixing in turn each one of the hadronic quantities to their physical value yields a physical value for the lattice spacing and for the gluonic scales. 
The results are tabulated in Table~\ref{tab:lattice_scales_hadronic}. 
\begin{table}
 \centering
 \begin{tabular*}{1.\textwidth}{@{\extracolsep{\fill}}rcccccc}
   \hline \hline
   & $a$ [fm] & $\sqrt{t_0}$ [fm]  & $(t_0/w_0)$ [fm]  & $w_0$ [fm]  & $r_0$ [fm] \\
   \hline
   $M_N$        & 0.0925(8)  & 0.1547(14) & 0.1395(13) & 0.1716(16) & 0.4913(63)\\
   $M_\pi$      & 0.0906(2)  & 0.1517(04) & 0.1367(03) & 0.1682(04) & 0.4816(45)\\
   $f_\pi$      & 0.0914(2)  & 0.1531(03) & 0.1380(03) & 0.1698(04) & 0.4861(45)\\
   $f_K$        & 0.0914(1)  & 0.1530(02) & 0.1380(02) & 0.1697(02) & 0.4860(44)\\
   \hline
   avg.         & 0.0913(2)   & 0.1528(03) &  0.1378(02) & 0.1695(03) & 0.4856(47) \\
   \hline \hline
 \end{tabular*}
 \caption{Lattice spacing and gluonic lattice scales in physical units determined from hadronic quantities at the physical point. The last line contains the weighted average of the four determinations assuming 100\% correlation.
   As input we have used $M_\pi = 134.98\ \mathrm{MeV}$, $f_\pi = 130.4(2)\ \mathrm{MeV}$, $f_K = 156.2(7)\ \mathrm{MeV}$ and the average proton-neutron mass $M_N = 938.9\ \mathrm{MeV}$, .
 }
 \label{tab:lattice_scales_hadronic}
\end{table}
In the last line we provide the weighted average, assuming 100\% correlation, together with the statistical error for the lattice spacing and the gluonic lattice scales determined from the hadronic quantities.
For the final error we also  include an estimate of the systematic error due to lattice artefacts and finite size effects. 
Our procedure to do so is the same as above leading to Eq.~\ref{eq:gluonic lattice spacing}.
We can allow for a possible small mismatch in the pion mass from its physical value and use lowest order (heavy baryon) chiral perturbation theory expression~\cite{Gasser:1987rb} to extrapolate to the physical point and then determine the lattice spacing. 
Applying this procedure for example to the nucleon mass~\footnote{For more details on this procedure we refer to Ref.~\cite{Abdel-Rehim:2015owa}.} we find a value of $a=0.093(1)$~fm consistent with the value obtained assuming that we are exactly at the physical point.

A weighted average of the lattice spacings, assuming 100\% correlation between our determinations, yields $a = 0.0913(2)$ fm with $\chi^2/\text{dof}=4.6$. 
The large value of $\chi^2/\text{dof}$ could simply be due to lattice artefact or due to the fact that our lattice data are not corrected for finite volume effects. 
In order to account for these uncertainties we quote a systematic error covering the spread of the determinations. 
In this way we obtain
\begin{equation}
  a_\mathrm{hadronic} = 0.0913(2)(11) \, \mathrm{fm} \, .
  \label{eq:hadronic lattice spacing}
\end{equation}
This lattice spacing is nicely consistent with the one determined from the gluonic quantities in Eq.~\ref{eq:gluonic lattice spacing} if the systematic error is taken into account. 
The weighted average of the two lattice spacings gives
\begin{equation}
  a = 0.0914(02)(15) \, \mathrm{fm} \, ,
  \label{eq:lattice spacing}
\end{equation}
where we quote the larger of the two estimates for the systematic error.

\begin{table}[t]
  \centering
  \footnotesize
  \begin{tabular*}{1.\textwidth}{@{\extracolsep{\fill}}rllll}
    \hline \hline \\[-2.0ex]
    & $f_K$                                 & $f_D$                              & $f_{D_s}$                         & $M_{D_s}$ \\[0.6ex]
    \hline \\[-2.0ex]
    ETM'09 [$\mev$]                 & $158.1(2.4)$                          & $197(9)$                           & $244(8)$                          & --  \\
    ETM'15 [$\mev$]                 & $155.0(1.9)$                          & $207.4(3.8)$                       & $247.2(4.1)$                      & --   \\
    PDG [$\mev$]                    & $156.2(7)$                            & $204.6(5.0)$                       & $257.5(4.6)$                      & $1968.50(32)$ \\
    FLAG [$\mev$]                   & $156.3(9)$                            & $209.2(3.3)$                       & $248.6(2.7)$                      & -- \\[0.6ex]
    \hline \hline
  \end{tabular*}
  \caption{Reference values for the dimensional quantities calcuated in this study. FLAG refers to $N_f=2+1$ averages from Ref.~\cite{2013:FlagReview}, ETM'09 refers to Ref.~\protect\cite{Blossier:2009bx} ($N_f=2$) and ETM'15 to Ref.~\protect{\cite{Carrasco:2014poa}} ($N_f=2+1+1$). 
  }
  \label{tab:pheno_lat_ps_values}
\end{table}

\subsection{Physical Scales}

In order to determine estimates of the gluonic scales $\sqrt{t_0},  t_0/w_0, w_0$ and $r_0$ at the physical point in physical units we proceed the same way as above. 
The weighted averages for $\sqrt{t_0},  t_0/w_0, w_0$ have a $\chi^2/\text{dof} \sim 4.5$, while the weighted average for $r_0$ has $\chi^2/\text{dof} =0.6$.
However, since our determinations of $aM_N, a\mpi, af_\pi$ and $af_K$ are certainly correlated, we do not expect a reduction of the error when averaging the four results. 
Assuming the data to be 100\% correlated, the error for $r_0$ for example increases from 0.0024 to 0.0047. 
In addition to the statistical error we again also quote a systematic error covering the spread of the determinations in order to account for the uncertainty due to lattice artefacts and/or finite volume effects. 
Eventually we obtain 
\begin{align*}
  \sqrt{t_0} &= 0.1528(03)(19)  \,\text{fm}\,,    \\
  t_0/w_0    &= 0.1378(02)(17)      \,\text{fm}\,,\\
  w_0        &= 0.1695(03)(19)    \,\text{fm}\,,  \\
  r_0        &=  0.4856(47)(57)   \,\text{fm}\,. 
\end{align*}
 
Finally, using the estimate of the lattice spacing from gluonic scales, hadronic quantities can be determined in physical units from their values listed in Table~\ref{tab:pions_afK_afD_afDs_aMDs}.
$f_K$, $f_D$, $f_{D_s}$ and $M_{D_s}$ take the values,
\begin{align*}
    f_K^\mathrm{(CD)}      &= 153.35  (0.18)(^{+0.04}_{-0.04})(2.96) ~\mev\,, \\
    f_D^\mathrm{(CD)}      &= 216.71  (1.99)(^{+0.59}_{-1.47})(4.19) ~\mev\,, \\ 
    f_{D_s}^\mathrm{(CD)}  &= 255.85  (0.49)(^{+0.10}_{-0.14})(4.95) ~\mev\,, \\ 
    M_{D_s}                &= 1912.3  (5.73)(^{+0.13}_{-0.15})(37.0) ~\mev\,,
\end{align*}
where the first error is statistical, the second is from the fit range ambiguity (see appendix~\ref{app:fitrange}) and the last one comes from the estimate of the lattice spacing.
They can be compared to pheomenological and lattice continuum limit results listed in Table~\ref{tab:pheno_lat_ps_values}.
The agreement to both FLAG and PDG is good when systematic errors are taken into account.
Likewise, the agreement to previous ETM continuum and chirally extrapolated results for $N_f=2$~\cite{Blossier:2009bx} and $N_f=2+1+1$~\cite{Carrasco:2014poa} is good.
We observe clearly that systematic errors are significantly bigger than the statistical ones.
Similarly, again using the estimate of the lattice spacing from gluonic scales, the nucleon mass can be given in physical units
\begin{equation*}
  M_N=933(8)(18)~\mev,
\end{equation*}
where the first error is statistical and the second stems from the estimate of the lattice spacing.
The agreement to the physical value of $M_N$ is excellent.

As an interesting experiment (assuming the absence of strange and charm quark effects and lattice artefacts in $M_N$), we can combine the $N_f=2+1+1$ data for $M_N$ at heavier than physical light quark masses with the new data at the physical point.
This procedure will allow to correct for the small missmatch in the physical pion mass value appearing when $M_N$ is used to set the scale.
To this end we consider SU(2) chiral perturbation theory ($\chi$PT) and the well-established $\mathcal{O}(p^3)$ result of the nucleon mass dependence on the pion mass~\cite{Gasser:1987rb,Bernard:1992qa}, given by
\begin{equation} 
  \label{eq:nucleon_p3}
  M_N = M_N^{0} - 4c_1\mpi^2 - \frac{3g_A^2}{32\pi \fpi^2} \mpi^3\,,
\end{equation}
where $M_N^0$ (the nucleon mass in the chiral limit) and $c_1$ are in general treated as fit parameters. 
In our fits, we fix the value of $c_1$ by constraining the fit to go through the physical point.
The lattice spacings $a_{\beta=1.90}$, $a_{\beta=1.95}$ and $a_{\beta=2.10}$ for the $N_f=2+1+1$ ensembles as well as the lattice spacing $a_\mathrm{phys}$ of our physical ensemble are considered as additional independent fit parameters.  
For the fit we find $\chi^2/\mathrm{dof}=1.578$ with $\mathrm{dof}=12$, corresponding to a $p$-value of $0.09$.
This procedure yields $a_\mathrm{phys}=0.093(1)$~fm, which is in agreement with the estimate of the lattice spacing in Eq.~\ref{eq:lattice spacing}, but shows some tension with the lattice spacing determined from other hadronic quantities as discussed above.
We remark that the lattice spacing determined from the nucleon mass is in very good agreement with the gluonic one, Eq.~\ref{eq:gluonic lattice spacing}.
In general, the discrepancy between the lattice spacing values determined from $M_N$ and $f_\pi$ is significantly reduced with the new action compared to the previous ETM $N_f=2$ and $N_f=2+1+1$ simulations.
 
\subsection{Quark Masses}\label{subsec:quarkmasses}

One advantage of having an ensemble with physical pion mass value available is the fact that the bare twisted mass can directly be related to the renormalised average up/down quark mass using appropriate renormalisation factors.
The relevant renormalisation factor in twisted mass lattice QCD is $Z_P$, since $m_q^\mathrm{R} = \mu_\ell / Z_P$.
We have determined $Z_P$ using the RI'-MOM scheme employing the momentum source technique~\cite{Gockeler:1998ye}.
Details are discussed in Refs.~\cite{Alexandrou:2015ren, Abdel-Rehim:2015owa}.
The value of $Z_P$ at $2\ \mathrm{GeV}$ in the $\overline{\mathrm{MS}}$ scheme reads
\[
Z_P = 0.501(8)(26)(12)\,.
\]
The first error is statistical, the second systematic stemming from the extrapolation to $(ap)^2 = 0$ and the perturbative subtraction of leading lattice artefacts~\cite{Constantinou:2009tr, Alexandrou:2010me} and the third from the conversion of RI' to $\overline{\mathrm{MS}}$ at $2\ \mathrm{GeV}$.

With $a\mu_\ell=0.0009$, the $Z_P$-value given above and the lattice spacing value from Eq.~\ref{eq:lattice spacing} we obtain a value for the average up/down quark mass as follows
\begin{equation}
  \label{eq:mlight}
  m_{ud}^{\overline{\mathrm{MS}}}(2\ \mathrm{GeV})\ = \ 3.88(6)(21)(10)\ \mathrm{MeV}\,,
\end{equation}
where the first error is statistical, the second from the combined systematic errors of the lattice scale and $Z_P$, and the third from the conversion of RI'-MOM to $\overline{\mathrm{MS}}$ at $2\ \mathrm{GeV}$.

Using the ratios $\mu_s/\mu_\ell$ and $\mu_c/\mu_\ell$ from Table~\ref{tab:mass_ratios} we can then directly compute also estimates for the strange and charm quark masses
\begin{equation}
  \label{eq:msc}
  \begin{split}
    m_s^{\overline{\mathrm{MS}}}(2\ \mathrm{GeV})\ &=\ 107(2)(6)(3)\ \mathrm{MeV}\,,\\
    m_c^{\overline{\mathrm{MS}}}(2\ \mathrm{GeV})\ &=\ 1.33(3)(7)(3)\ \mathrm{GeV}\,.
  \end{split}
\end{equation}
Both, $m_{ud}$ and $m_s$ compare well to the quark mass values determined on the $N_f=2$ ETMC ensembles without clover term~\cite{Blossier:2007vv}. 
We can also compare $m_{ud}$ and $m_s$ to the $N_f=2$ determinations from Refs.~\cite{Blossier:2010cr, Bazavov:2010yq, Arthur:2012opa, Durr:2010vn, Durr:2010aw} averaged by FLAG~\cite{Aoki:2013ldr}, namely
\[
m_{ud}=3.6(2)\ \mev\,,\qquad m_s=101(3)\ \mev\,,
\]
which are both in agreement with our determinations.
The values presented above should be taken with some care, because we did not take the continuum and thermodynamic limits.
An alternative determination of $Z_P$~\cite{Constantinou:2010gr, Cichy:2012is} might shed light on the fact that all three quark masses determined here have consistently larger values than what can be found in the literature, while the quark mass ratios show good agreement.

\subsection{Lepton Anomalous Magnetic Moments}

\begin{figure}[t]
  \centering
  \includegraphics[width=0.8\textwidth]{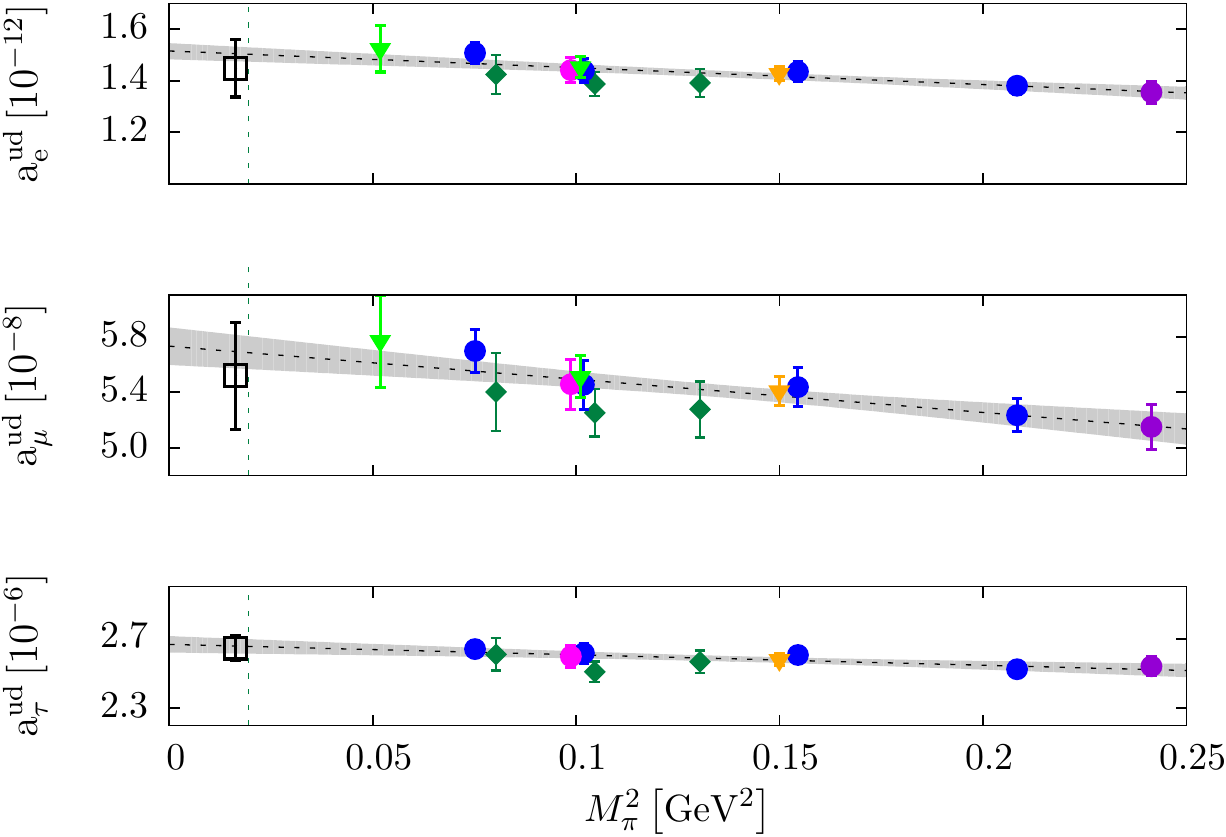}
  \caption{Comparison of the chiral extrapolation of the light quark contributions to the three lepton anomalous magnetic moments obtained from $N_f=2+1+1$ simulations to the values 
    obtained with the standard definition 
    Eq.~\protect\ref{eq:amudef} at the physical value of the 
    pion mass (black square). 
    The dark green diamonds correspond to
    $a=0.086\mathrm{ fm}$ and $L=2.8\mathrm{ fm}$ and the circles to $a=0.078\mathrm{ fm}$, the violet one stands for $L=1.9\mathrm{ fm}$, the blue ones for $L=2.5\mathrm{
      fm}$, and the pink for $L=3.7\mathrm{ fm}$. The orange triangle shows the value obtained for $a=0.061\mathrm{ fm}$ and $L=1.9\mathrm{ fm}$ and the light green
    triangle denotes $a=0.061\mathrm{ fm}$ and $L=2.9\mathrm{ fm}$.}
  \label{fig:gm2_light_all}
\end{figure}

In this section, we discuss the leading-order light quark hadronic contribution to the anomalous magnetic moments of the electron, the muon and the $\tau$ leptons, $a_e^\mathrm{ud}$, $a_\mu^\mathrm{ud}$ and $a_\tau^\mathrm{ud}$, respectively. 
We have performed exactly the same analysis as described in Ref.~\cite{Burger:2013jya} for the anomalous magnetic moment of the muon, only changing the lepton masses in the numerical integration to the ones of the electron and $\tau$ lepton.
We will compare the results obtained at the physical point with the ones that were obtained from ensembles at unphysically large pion masses and which were then extrapolated to the physical point. 
In Figure~\ref{fig:gm2_light_all} we show the data for the three $a^\mathrm{ud}$ as a function of $\mpi^2$ comparing the results of Refs.~\cite{Burger:2013jya, Burger:2015oya} with the new result at the physical point.

For our results at unphysically large pion masses we have used the same redefinition of the vacuum polarisation function as in Refs.~\cite{Feng:2011zk, Renner:2012fa, Burger:2013jya, Burger:2015oya}
\begin{equation}
  \label{eq:redef}
  a_{\overline{\mathrm{l}}}^{\mathrm{hvp}} = \alpha^2 \int_0^{\infty} \frac{d Q^2 }{Q^2} w\left( \frac{Q^2}{H^2}
    \frac{H_{\mathrm{phys}}^2}{m_l^2}\right) \Pi_{\mathrm{R}}(Q^2) \,,
\end{equation}
with the hadronic scale $H=M_V$, the lowest lying vector meson state, and $m_l$ the lepton mass. 
$H=H_\mathrm{ phys}=1$ corresponds to the standard definition given in Eq.~\ref{eq:amudef}.

When determining the lepton anomalous magnetic moments the chiral extrapolation to the physical pion mass can lead to a severe systematic error. 
This uncertainty is avoided when using ensembles at the physical point~\cite{Abdel-Rehim:2013yaa}. 
We have computed the light quark contributions to the lepton anomalous magnetic moments with the standard definition Eq.~\ref{eq:amudef} on 800 configurations of the new physical ensemble. 
We find full agreement with our previous results for the light quark contribution originating from a chiral extrapolation of our $N_f=2$ as well as $N_f=2+1+1$ results. 
The extrapolations of the $N_f=2+1+1$ data are depicted in Figure~\ref{fig:gm2_light_all} as dashed lines with shaded error band, whereas the extrapolated values -- also including the previous $N_f=2$ values from Ref.~\cite{Feng:2011zk} -- are given in Table~\ref{tab:results_gm2_light}. 

\begin{table}[htb]
  \begin{center}
    \begin{tabular*}{1.\textwidth}{@{\extracolsep{\fill}}crrr}
      \hline \hline
      &  physical point & extr. $N_f=2$ & extr. $N_f=2+1+1$ \\
      \hline \hline
      & & & \vspace{-0.40cm} \\
      $a_\mathrm{ e }^\mathrm{ hvp}\cdot 10^{12}$ & $1.45(11)$ & $1.51(04)$ & $1.50(03)$  \\
      $a_\mathrm{ \mu}^\mathrm{ hvp}\cdot 10^{8}$ & $5.52(39)$ & $5.72(16)$ & $5.67(11)$\\
      $a_\mathrm{ \tau}^\mathrm{ hvp}\cdot 10^{6}$ & $2.65(07)$ & $2.65(02)$ & $2.66(02)$ \\
      \hline \hline
    \end{tabular*}
    \caption{\label{tab:results_gm2_light} Comparison of the values for $a_{\mathrm{e}}^\mathrm{ hvp}$, $a_{\mathrm{\mu}}^{\mathrm{hvp}}$, and
      $a_{\mathrm{\tau}}^{\mathrm{hvp}}$ obtained at the physical point using the standard definition Eq.~\protect\ref{eq:amudef} with the
      results of the linear extrapolations from our improved definition Eq.~\protect\ref{eq:redef} on the $N_f=2$
      and $N_f=2+1+1$ ETMC ensembles without clover term.} 
  \end{center}
\end{table}

We made a particular effort to quantify the systematic uncertainties which arise in our calculation for the lepton anomalous magnetic moments in the data not at the physical point. 
These systematic effects originate from the chiral extrapolation, the continuum limit, the fit range for the vector meson mass and the form of the fit function. 
These investigations are described in detail in Ref.~\cite{Burger:2015oya}. 
We also compared the approach of using Pad\'e fits as proposed in Ref.~\cite{Aubin:2012me} with our MNBC fits. 
As discussed in Ref.~\cite{Burger:2013jva} we could not find a clear advantage of using the Pad{\'e} approach which led us therefore to stay with our standard fit function.

\section{Summary and Discussion}
\label{sec:summary}
In this paper we have presented results from a first simulation with two flavours of Wilson twisted mass fermions at maximal twist, directly at the physical value of the pion mass.
The introduction of the clover term in the action significantly reduces the mass splitting between neutral and charged pions, which previously prevented such simulations to be carried out.
As we have demonstrated in this paper, with the clover parameter close to its non-perturbatively tuned value, simulations at the physical point are stable even at a lattice spacing of $a \approx 0.09\ \mathrm{fm}$ with no signs of meta-stabilities and the pion mass splitting vanishes within our uncertainties.
Thus, the action discussed here can be used also for smaller values of the lattice spacing such that eventually a continuum limit extrapolation of lattice results can be performed.
We consider the present work as a first step in this direction. 

It must be stressed that with the addition of the clover term, the arguments for {\em automatic} $\mathcal{O}(a)$ improvement at maximal twist for twisted mass lattice QCD hold at any value of the clover parameter.
Therefore, as in the past, only the Wilson quark mass needs to be tuned to achieve the automatic $\mathcal{O}(a)$ improvement. 
As a consequence, with the action used here, all physical quantities considered in the broad research program of our collaboration scale with a rate of $\mathcal{O}(a^2)$ to the continuum limit and no additional operator specific improvement coefficients are needed.
We regard this fact as a major advantage of the maximally twisted mass approach to compute physical quantities from lattice QCD.

On the ensemble at the physical pion mass we have computed phenomenologically interesting observables including $f_K, f_D, f_{D_s}$, quark masses and ratios thereof, the nucleon mass and the hadronic contribution to the anomalous magnetic moments of leptons. 
Where possible, we compared to previous results obtained with twisted mass fermions at unphysically large pion mass values with $N_f=2$ and $N_f=2+1+1$ dynamical quark flavours without clover term.
First results on meson and nucleon structure using this new ensemble are presented in Ref.~\cite{Abdel-Rehim:2015owa}.
The physical pion mass ensemble with clover term largely confirms the reliability of the chiral extrapolations performed with ETMC ensembles without clover term. Note, however, that we currently have only one value of the lattice spacing available with the new action and only one volume at the physical point. 
This means that we cannot (directly) control finite volume and finite lattice spacing effects and, hence, the systematic errors might be larger than our current estimates.
For the lattice artefacts we have an indirect way of estimating them using ensembles generated with the new action but with larger than physical pion mass values.
These can be compared to ETMC's $N_f=2$ simulations without clover term at similar volumes and pion masses, indicating that with the clover term in the action, $\mathcal{O}(a^2)$ lattice artefacts are small.
However, not unexpectedly, lattice artefacts become visible for quantities in the heavy quark sector. 

\begin{figure}[b]
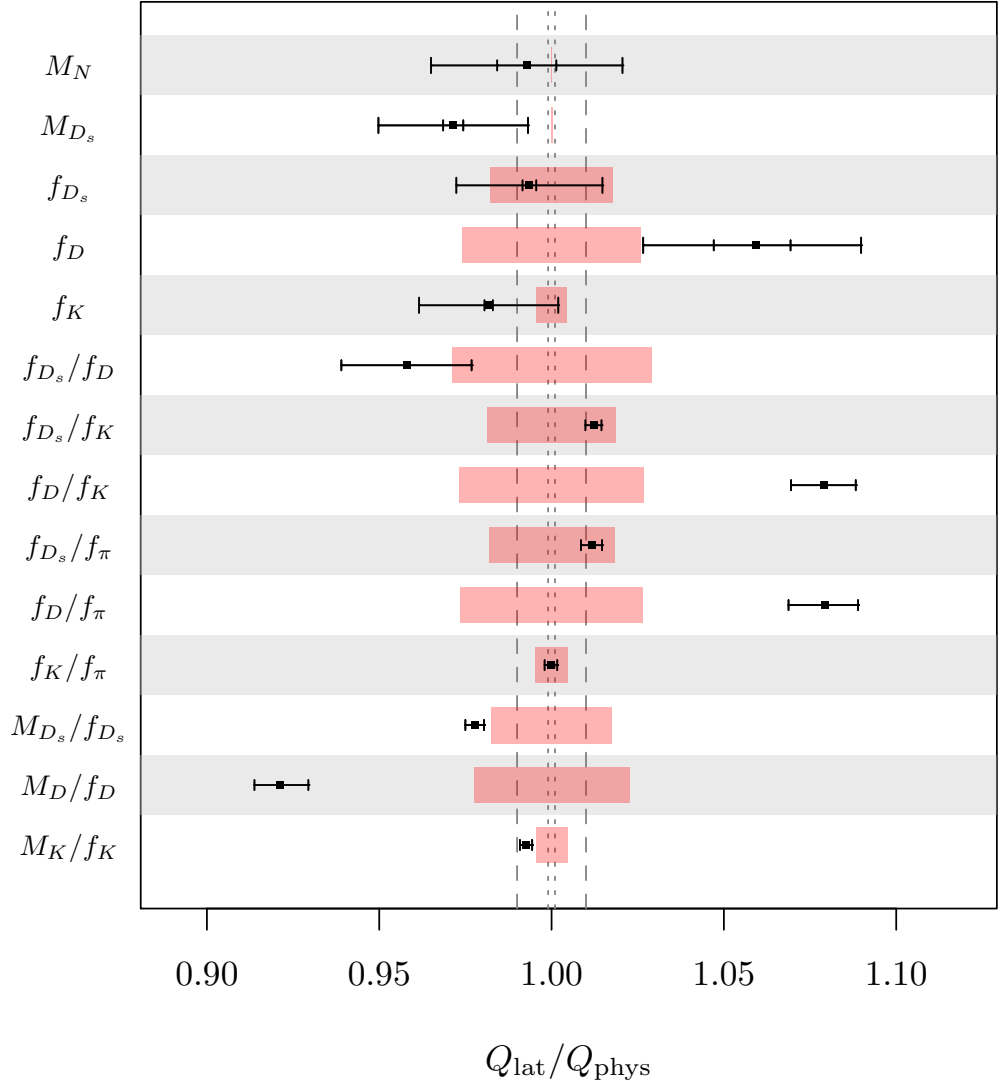

 \centering
 \includegraphics[width=0.8\linewidth]{{{Qlat_ov_Qphys.linear}}}
 \caption{Ratios of lattice results and phenomenological values of the quantities in the legend with lattice decay constants computed via the continuum definition. For dimensional quantities, the inner error bar combines the statistical and systematic errors in quadrature while the outer error bar stems from the estimate of the lattice spacing from gluonic scales. The red bands show the phenomenological uncertainty on $Q_\mathrm{phys}$ separately (the respective experimental errors on $M_N$ and $M_{D_s}$ are too small to be visible). The dotted and dashed lines indicate per-mille and per-cent deviations from $1.0$ respectively.}
 \label{fig:Qlat_ov_Qphys}
\end{figure}

We have determined the lattice spacing using gluonic and hadronic observables. 
We found good agreement between gluonic and hadronic determinations of the lattice spacing after averaging over different scale setting quantities.
When comparing the scales obtained by considering observables such as $\fpi$ or $\mpi$ in isolation, the corresponding differences are not covered by the statistical errors. 
This is of course not unexpected and it is likely that lattice artefacts are mainly responsible for these differences, which we account for through the quoted systematic error.

Given this estimate of the lattice spacing, we are able to predict values for other, independent quantities in physical units.
Figure~\ref{fig:Qlat_ov_Qphys} shows a comparison between the values of various quantities $Q$ determined in the present analysis and their phenomenological counterparts in the form $Q_\mathrm{lat}/Q_\mathrm{phys}$.
It is notable that for most quantities the current analysis gives error estimates that are of the same order or smaller than their phenomenological ones.
We observe -- with the exception of a few $D$ and $D_s$ meson related quantities -- agreement within errors between lattice and phenomenological determinations.
Lattice artefacts are probably responsible for the deviations and in the case of $D$-meson quantities, cut-off effects below 10\% can be considered as tolerably small.

A number of conclusions can be drawn concerning the effect of simulations at the physical point on systematic errors.
The most striking feature is certainly the potential precision that quark mass ratios can be determined with, as indicated by the rather small errors given in Table~\ref{tab:mass_ratios}.
Together with the lattice spacing estimates and the computation of the renormalisation constant, this allows us to determine the average up/down quark mass and the strange and charm quark masses without any chiral extrapolation.
The agreement with other $N_f=2$ determinations is good.

In the baryon sector, employing chiral effective theory for the quark mass extrapolation of quantities like $g_A$ of the nucleon typically leads to large uncertainties.
Here the calculations directly at the physical point are, therefore, very helpful to give insights on the source of the discrepancies of specific observables between lattice QCD and phenomenological results, see Ref.~\cite{Abdel-Rehim:2015owa}. 
For lepton anomalous magnetic moments, simulations at the physical value of the light quark mass confirm the correctness of the lattice redefinition of the $Q^2$ dependence of the weight function within errors.
Here, an open problem is the allowed strong decay of the $\rho$ meson, which has to be taken into account in the computations of the light quark contribution to the hadronic vacuum polarisation and thus the lepton moments $a_l, l \in \lbrace e,\mu,\tau \rbrace$ and the electroweak couplings.

Currently we are generating an ensemble with the new action and physical pion mass value at the same lattice spacing value but with $L/a=64$.
This ensemble, once completed, will allow us to address finite volume effects for the quantities discussed here and the nucleon observables presented in Ref.~\cite{Abdel-Rehim:2015owa} in the near future.
We take the results of this paper as indicative of the fact that simulations at the physical pion mass with the symmetry-based $\mathcal{O}(a)$-improvement of maximally twisted mass quarks will allow many interesting quantities to be accureately studied. 
In particular, we expect that we will eventually be able to provide theoretical input with competitive uncertainties relevant to the analysis of experimental data for heavy flavour physics, hadron scattering and lattice studies of resonances and the baryon sector, as well as tests of the Standard Model in the electroweak sector.
This is especially true of our ongoing simulations using $N_f=2+1+1$ flavours of clover-improved twisted mass quarks at the physical point.

\acknowledgments
We would like to thank all members of ETMC for the most enjoyable collaboration.
B.K. gratefully acknowledges full financial support under AFR PhD grant 2773315 of the National Research Fund, Luxembourg.
We acknowledge partial financial support from the Deutsche Forschungsgemeinschaft (Sino-German CRC 110).
K. H.  and M. C. are partly supported by  the Cyprus Research Promotion Foundation under contract T$\Pi$E/
$\Pi\Lambda$HPO/ 0311(BIE)/09 and TECHNOLOGY/ $\Theta$E$\Pi$I$\Sigma$/ 0308(BE)/17, respectively.
The computer time for this project was made available to us by the John von
Neumann-Institute for Computing (NIC) on the Juqueen system in J{\"u}lich, as well as on systems of the Swiss National Supercomputing Centre (CSCS) under project ID s540 and HPC resources of CINES and IDRIS under the
allocation 52271 made by GENCI.
The open source software packages tmLQCD~\cite{Jansen:2009xp, Abdel-Rehim:2013wba, Deuzeman:2013xaa}, Lemon~\cite{Deuzeman:2011wz} and R~\cite{R:2005} have been used.
 
\appendix
\section{Simulation Details}\label{app:simdetails}

\subsection{Tuning to Maximal Twist}

\begin{figure}[h]
 \centering
 \includegraphics[width=0.9\linewidth]{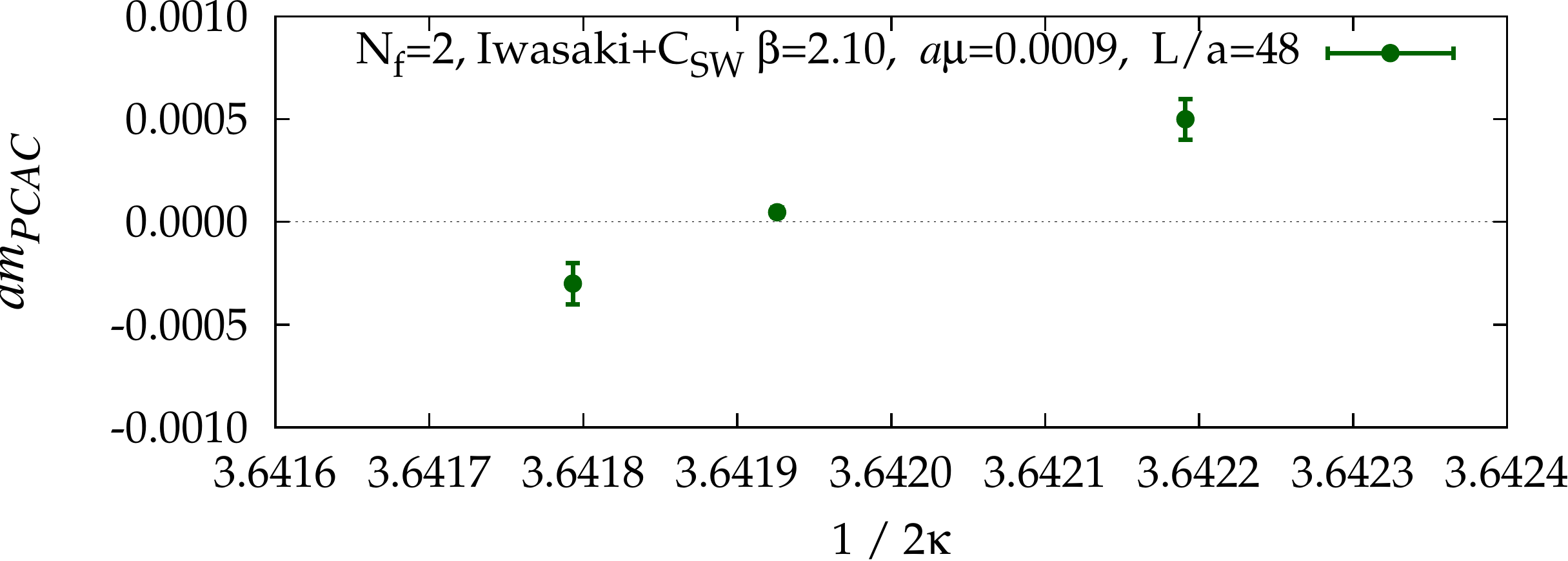}
 \caption{$am_{\mathrm{PCAC}}$ as a function of $1/2\kappa$ for the
   tuning of ensembles \textit{cA2.09.48} with $\kappa = (2am_0 + 8)^{-1}$.}
 \label{fig:mpcac_kappa_nf2phys}
\end{figure}

For the purpose of automatic $\mathcal{O}(a)$-improvement an estimate of the critical value of the hopping parameter can be determined from Ref. \cite{Aoki:2005et} with a relatively large error.
This value was refined through two simulations and a linear interpolation in $1/2\kappa$ as shown in Figure~\ref{fig:mpcac_kappa_nf2phys}.
In past simulations employing twisted mass fermions without a clover term, it was seen that the linear behaviour shown in Figure~\ref{fig:mpcac_kappa_nf2phys} breaks down around the critical $\kappa$ value, making interpolations difficult.
In addition, a much steeper slope than shown here forced us to perform very fine adjustments of the hopping parameter, which in turn required long tuning runs to control the statistical error on the individual measurements.
As was discussed in Ref.~\cite{Boucaud:2008xu} and shown in subsequent simulation results, it is sufficient to tune the renormalised PCAC quark mass to be no larger than 10\% of the renormalised twisted quark mass to ensure $\mathcal{O}(a)$ improvement in practice.
For the \textit{cA2.09.48} ensemble, the \textit{bare} PCAC quark mass takes a value of $a\mpcac \approx 8(1)\cdot10^{-5}$, thus fulfilling the condition.

\subsection{Molecular Dynamics Histories}

\begin{figure}
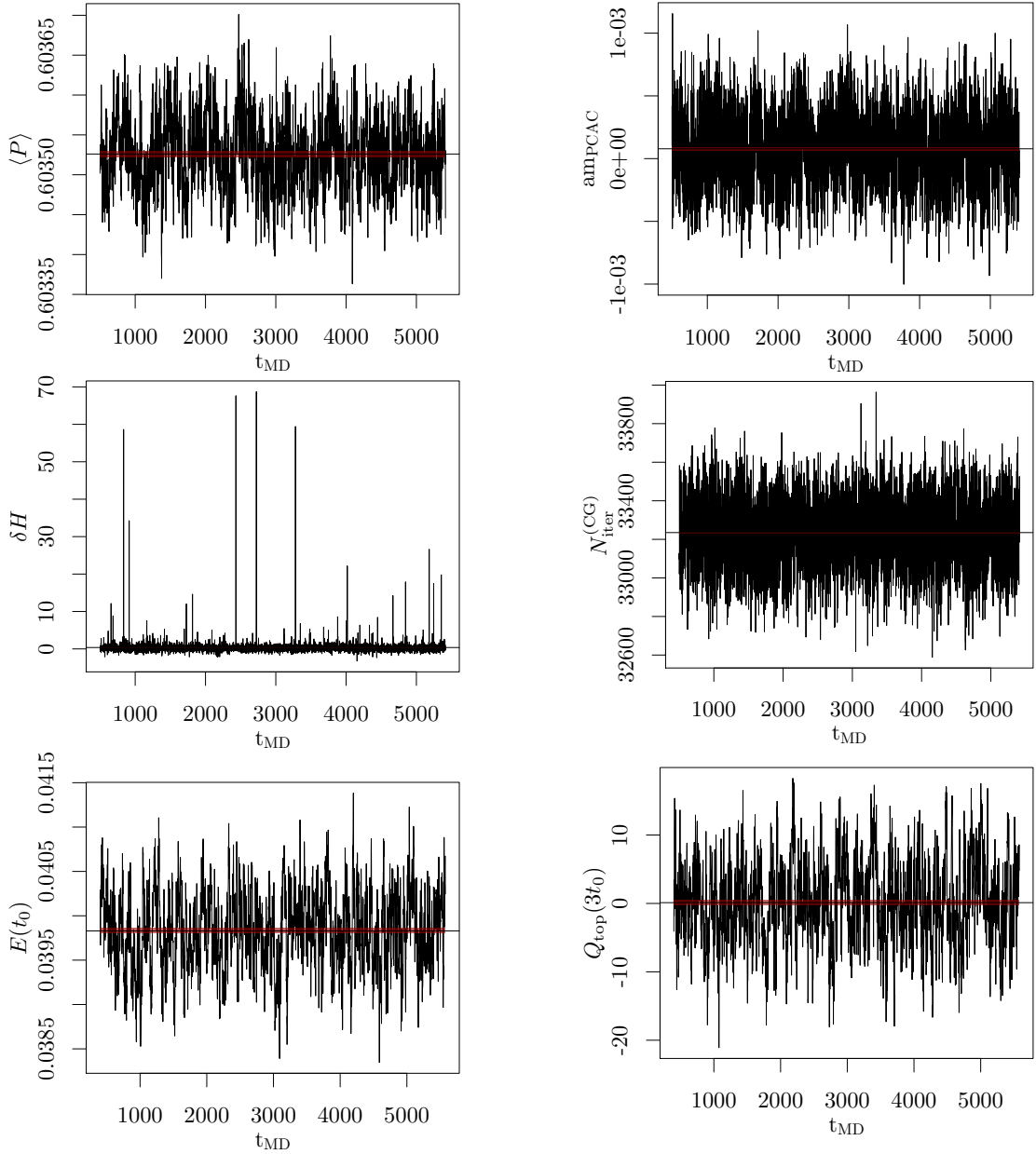

 \begin{subfigure}{0.49\textwidth}
  \centering
  \includegraphics[page=7,width=0.8\linewidth]{{{md_histories_iwa_b2.1-L48T96-csw1.57551-k0.13729-mul0.0009}}}
 \end{subfigure}
 \begin{subfigure}{0.49\textwidth}
  \centering
  \includegraphics[page=1,width=0.8\linewidth]{{{md_histories_iwa_b2.1-L48T96-csw1.57551-k0.13729-mul0.0009}}}
 \end{subfigure}
 \begin{subfigure}{0.49\textwidth}
  \centering
  \includegraphics[page=11,width=0.8\linewidth]{{{md_histories_iwa_b2.1-L48T96-csw1.57551-k0.13729-mul0.0009}}}
 \end{subfigure}
 \begin{subfigure}{0.49\textwidth}
  \centering
  \includegraphics[page=19,width=0.8\linewidth]{{{md_histories_iwa_b2.1-L48T96-csw1.57551-k0.13729-mul0.0009}}}
 \end{subfigure}
 \begin{subfigure}{0.49\textwidth}
  \centering
    \includegraphics[page=1,width=0.8\linewidth]{{{md.history.Esymt0}}}
 \end{subfigure}
 \begin{subfigure}{0.49\textwidth}
  \centering
    \includegraphics[page=1,width=0.8\linewidth]{{{md.history.Qtop3t0}}}
 \end{subfigure}
  \caption{Molecular dynamics histories of various quantities on ensemble \textit{cA2.09.48}. }
  \label{fig:md_histories}
\end{figure}

As discussed in section~\ref{sec:introduction}, in past simulations, lattice artefacts rendered simulations without a clover term meta-stable as the pion mass was lowered towards its physical value.
As can be seen in Figure~\ref{fig:md_histories}, the twisted mass clover action results in very stable molecular dynamics histories at the physical average up/down quark mass without any signs of meta-stability in the plaquette or the PCAC quark mass despite the relatively coarse lattice spacing in excess of $0.09~\mathrm{fm}$.
As expected, the topological charge in the Wilson flow definition and the energy density at flow time $t_0$ as defined in section~\ref{subsec:gluonic scales} are sampled well and show integrated autocorrelation times well below one hundredth of the total simulation time.
This can be seen in the bottom-most panels of Figure~\ref{fig:md_histories}.

A complete listing of various algorithmic observables from the different ensembles is provided in Table~\ref{tab:online measurements} together with their autocorrelation times.
For ensemble \textit{cA2.30.24}, it is to be notedz that there are significant fluctuations in the number of CG iterations, probably an indication of an insufficient volume for the simulated pion mass.

\begin{table}
 \centering
 \small
 \begin{tabular*}{1.\textwidth}{@{\extracolsep{\fill}}rllll}
  \hline \hline \\[-2.0ex]
  observable                                                       & \textit{cA2.09.48} & \textit{cA2.30.24} & \textit{cA2.60.24} & \textit{cA2.60.32} \\[0.6ex]
  \hline \hline \\[-2.0ex]
  $P_\mathrm{acc}$                                                 & 0.726(6)           & 0.910(7)           & 0.771(5)           & 0.874(4)  \\
  $\langle P \rangle$                                              & 0.603526(4)        & 0.603562(9)        & 0.603535(5)        & 0.603533(2) \\
  $\langle \mpcac \rangle$                                         & 0.00008(1)         & -0.00037(7)        & -0.00026(3)        & -0.00021(1) \\  
  $\langle \delta H \rangle$                                       & 0.37(3)            & 0.047(12)          & 0.177(8)           & 0.044(3) \\
  $\langle \exp(-\delta H) \rangle$                                 & 1.00(1)            & 1.01(1)            & 1.003(7)           & 1.003(3) \\
  $\langle N_\mathrm{iter}^\mathrm{(CG)} \rangle$                  & 33235(3)           & 10720(67)          & 5288(2)            & 5674(1) \\[1.0ex]   
  $\tau_\mathrm{int}\lbrace P \rbrace$                             & 15(5)              & 3.2(8)             & 3.8(7)             & 2.9(5) \\
  $\tau_\mathrm{int}\lbrace \mpcac \rbrace$                        & 15(5)              & 1.6(4)             & 1.4(2)             & 1.2(1) \\
  $\tau_\mathrm{int}\lbrace\delta H\rbrace$                        & 0.50(4)            & 0.50(2)            & 0.53(3)            & 0.50(1) \\  
  $\tau_\mathrm{int}\lbrace\exp(-\delta H)\rbrace$                  & 0.49(2)            & 0.48(2)            & 0.49(1)            & 0.50(1) \\
  $\tau_\mathrm{int}\lbrace N_\mathrm{iter}^\mathrm{(CG)}\rbrace$  & 0.83(9)            & 17(9)              & 1.8(2)             & 4.2(8) \\[0.6ex]
  \hline \hline
 \end{tabular*}
 \caption{Expectation values and autocorrelation times of various observables for ensembles used in this study. $N_\mathrm{iter}^\mathrm{(CG)}$ refers to the number of CG iterations in the heat-bath and acceptance steps of the mass preconditioning determinant ratio which has the target light quark mass in the numerator. $P_\mathrm{acc}$ refers to the acceptance rate which should be used to scale the autocorrelation times, which are given in units of trajectories.}
 \label{tab:online measurements}
\end{table}

A feature of the simulation which deserves a special mention is the behaviour of the energy violation which we denote by $\delta H$.
It seems that compared to twisted mass simulations without the clover term, large deviations occur quite frequently but they do not seem to affect the stability of the algorithm, do not seem to affect any observables and are in line with what has been observed by other collaborations~\cite{Durr:2010aw, Ukita:2008mq}.

\subsection{Simulation Parameters}

\begin{figure}
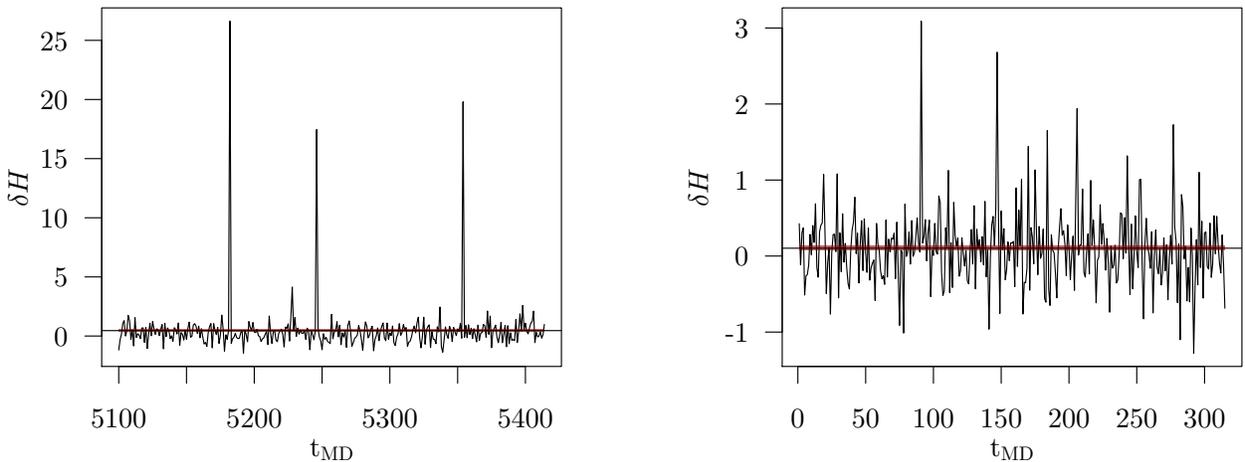

 \centering
 \begin{subfigure}{0.45\linewidth}
  \includegraphics[width=\linewidth,page=11]{{{md_histories_cA2.09.48.detail}}}
 \end{subfigure}
 \hfill
 \begin{subfigure}{0.45\linewidth}
  \includegraphics[width=\linewidth,page=11]{{{md_histories_cA2z.09.48.detail}}}
 \end{subfigure}
 \caption{Detail of MD histories of the energy violation $\delta H$ for runs \textit{cA2.09.48}(left) and \textit{cA2z.09.48}(right), indicating a reduction of large deviations as the number of integration steps is increased and the time scale splitting is adjusted.}
 \label{fig:md_histories_detail}
\end{figure}

The simulation parameters for the ensembles used in this work are listed in Table~\ref{tab:sim_params}, including the mass preconditioning and the number of integration steps on the various timescales.
In order to determine the origin of the sizeable $\delta H$ fluctuations observed in the molecular dynamics history of \textit{cA2.09.48}, short simulations \textit{cA2x.09.48}, \textit{cA2y.09.48} and \textit{cA2z.09.48} with more integration steps and more timescales were performed.
As shown in Figure~\ref{fig:md_histories_detail} it was found that this significantly reduces the frequency of large energy violations at the price of increased simulation cost.
It should be noted that none of the observables that we determined on ensemble \textit{cA2z.09.48} showed any deviation within errors compared to those on ensemble \textit{cA2.09.48}.

\begin{table}
 \centering
 \footnotesize
 \begin{tabular*}{1.\textwidth}{@{\extracolsep{\fill}}rlll}
  \hline \hline \\[-1.8ex]
  ensemble            & $N_t$            & $a\rho_t^\mathrm{HB}$                                                                         & $\log_{10}\left({^{r^2_a}_{r^2_f}}\right)$                                          \\[1.0ex]
  \hline \\[-1.8ex]
  \textit{cA2.60.24}  & $\{1,2,2,7\}$    & $\{-,0.060,{^{0.0110}_{0.0600}},{^{0.0000}_{0.0110}}\}$                                       & $\{-,{^{-22}_{-14}},{^{-22}_{-14}},{^{-22}_{-14}}\}$                                \\[0.6ex]
  \textit{cA2.60.32}  & $\{1,1,1,1,14\}$ & $\{-,0.800,{^{0.0800}_{0.8000}},{^{0.0080}_{0.0800}},{^{0.0000}_{0.0080}}\}$                  & $\{-,{^{-22}_{-14}},{^{-22}_{-14}},{^{-22}_{-14}},{^{-22}_{-14}}\}$                 \\[0.6ex]
  \textit{cA2.30.24}  & $\{1,2,2,10\}$   & $\{-,0.040,{^{0.0080}_{0.0400}},{^{0.0000}_{0.0080}}\}$                                       & $\{-,{^{-22}_{-14}},{^{-22}_{-14}},{^{-22}_{-14}}\}$                                \\[0.6ex]
  \textit{cA2.09.48}  & $\{1,1,2,13\}$   & $\{-,0.030,{^{0.0050}_{0.0300}},\left[ {^{0.0013}_{0.0050}},{^{0.0000}_{0.0013}} \right] \}$  & $\{-,{^{-22}_{-14}},{^{-22}_{-14}},\left[ {^{-22}_{-14}},{^{-22}_{-14}} \right] \}$ \\[0.6ex]
  \textit{cA2x.09.48} & $\{1,1,2,17\}$   & $\{-,0.030,{^{0.0050}_{0.0300}},\left[ {^{0.0013}_{0.0050}},{^{0.0000}_{0.0013}} \right] \}$  & $\{-,{^{-22}_{-14}},{^{-22}_{-14}},\left[ {^{-22}_{-14}},{^{-22}_{-14}} \right] \}$ \\[0.6ex]
  \textit{cA2y.09.48} & $\{1,1,1,1,13\}$ & $\{-,0.250,{^{0.0250}_{0.2500}},{^{0.0025}_{0.0250}},{^{0.0000}_{0.0025}}\}$                  & $\{-,{^{-22}_{-14}},{^{-22}_{-14}}, {^{-22}_{-14}}, {^{-22}_{-14}} \}$              \\[0.6ex]
  \textit{cA2z.09.48} & $\{1,1,1,1,17\}$ & $\{-,0.250,{^{0.0250}_{0.2500}},{^{0.0025}_{0.0250}},{^{0.0000}_{0.0025}}\}$                  & $\{-,{^{-22}_{-14}},{^{-22}_{-14}}, {^{-22}_{-14}}, {^{-22}_{-14}} \}$              \\[0.6ex]
  \hline \hline
 \end{tabular*}
 \caption{Simulation parameters for the ensembles used in this work and three additional test ensembles. $N_t$: number of integration steps of second order minimal norm integrator on the various time-scales. $a\rho_t^\mathrm{HB}$: Hasenbusch mass pre-conditioning parameters as in Ref.~\cite{Hasenbusch:2002ai} but with multiple determinant ratios. $r^2_a (r^2_f)$: squared relative residual stopping criterion in the acceptance step (force calculation) in the conjugate gradients solver. Square brackets indicate that more than one monomial is placed on the same timescale.}
 \label{tab:sim_params}
\end{table}

 \section{Pseudo-scalar Meson Analysis Details}\label{app:meson_details}

In this section we discuss the methods adopted for the analysis of pseudo-scalar meson correlators to produce the results of section~\ref{subsec:meson_results}.
First we introduce a somewhat novel technique for quantifying the uncertainty due to the choice of fit range for correlation functions.
Then we discuss different methods for choosing the bare valence strange and charm quark masses and how this choice affects the central values and uncertainties on the physical quantities extracted from the analysis.
Finally we show some examples of linear interpolations of the various quantities presented in this analysis and discuss remaining uncertainties such as discretisation artefacts and finite-volume corrections which have not been accounted for yet.

For each gauge configuration, the quark propagators for all masses were computed from the same stochastic ($Z_2$) time-slice sources and the correlation functions were constructed using the one-end trick.
A single time-slice source, chosen at random, with full spin-dilution was used for each gauge configuration and local-local, fuzzed-local and fuzzed-fuzzed correlation functions were computed to improve the extraction of the ground state mass in a constrained $2\times2$ matrix fit.
To profit from correlations in the data, the complete analysis was carried out in a stationary blocked bootstrap~\cite{stationarybootstrap} framework with block lengths tuned to accommodate the short autocorrelations in the data as determined from the Gamma method.
All observables were bootstrapped with the same bootstrap samples, preserving all correlations.

Firstly, positive correlation reduces the statistical error in many expressions built from data, in particular in ratios.
Secondly, preserving correlation at all levels in the analysis allows one to provide realistic error estimates on our final results.

\subsection{Computing the Neutral Pion Mass}\label{app:neutralpion}

\begin{figure}
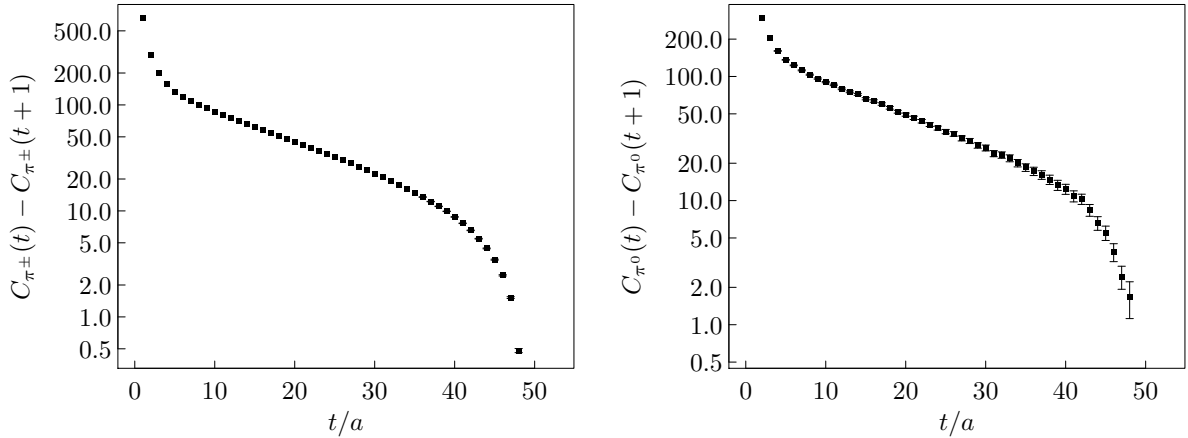

 \centering
 \includegraphics[width=0.46\textwidth,page=1]{{{cA2.09.48.pi0.corrs}}} \hspace{0.3cm}
 \includegraphics[width=0.46\textwidth,page=2]{{{cA2.09.48.pi0.corrs}}}
 \caption{Shifted two-point correlation functions (see Eq.\ref{eq:shifted_pi_corr}) for the charged (left) and full neutral (right) pion computed on ensemble \emph{cA2.09.48} for $\delta t = 1$.}
 \label{fig:shifted_pi_corr}
\end{figure}

The neutral pion two-point correlation function in the twisted mass formulation is notoriously difficult to compute with sufficient statistical precision, because it requires the inclusion of quark-line disconnected diagrams.
In addition, it is contaminated by a very noisy vacuum expectation value~\cite{Jansen:2005cg} because in tmLQCD, the neutral pion has the same quantum numbers as the vacuum state.
In previous analyses, see e.g. Refs.~\cite{Jansen:2005cg,Herdoiza:2013sla}, this contamination was explicitly computed and subtracted from the disconnected piece of the correlation function, which may result in biases if the number of measurements is insufficient to properly estimate the ensemble average of the vacuum expectation value.
We have adopted new techniques to efficiently deal with both issues.

In order tackle the first difficulty, we construct the neutral pion correlation functions using the stochastic Laplacian Heavyside method~\cite{Peardon:2009gh,Morningstar:2011ka} in the context of Ref.~\cite{Liu:2016cba}.
In this setup, between 3000 and 4000 inversions per gauge configuration are carried out for the computation of so-called perambulators, which can be stored efficiently and later used to construct any correlation function with the given quark content, including statistically well-resolved quark-line disconnected contributions. 
For details, see Ref.~\cite{Helmes:2015gla}.

\begin{figure}
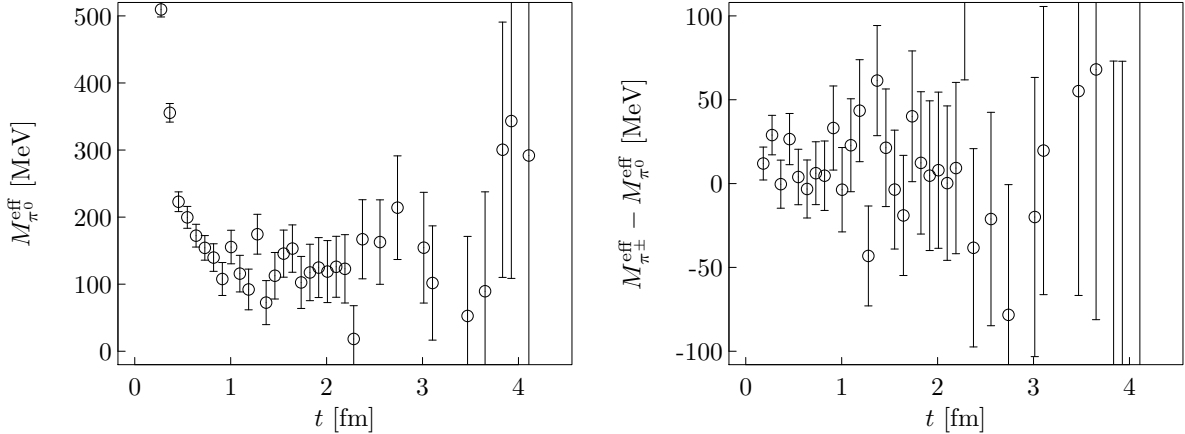

 \centering
 \includegraphics[width=0.46\textwidth,page=4]{{{cA2.09.48.pi0.corrs.effmasses}}} \hspace{0.3cm}
 \includegraphics[width=0.46\textwidth,page=5]{{{cA2.09.48.pi0.corrs.effmasses}}}
 \caption{(left) Effective mass for the full neutral pion, obtained by solving Eq.\ref{eq:shifted_effmass} for each source sink separation with $\delta t = 1$. (right) Correlated difference of the effective masses from the charged and full neutral pion correlation functions with $\delta t = 1$.}
 \label{fig:shifted_pi_effmasses}
\end{figure}

The second issue is approached instead by observing that the vacuum expectation value contributes the same constant to the two-point function at every source-sink separation.
In order to remove this contribution on a configuration by configuration basis (rather than subtracting the ensemble average of the vacuum expectation value), we consider the difference of the correlation function at two neighbouring source sink separations.
Thus, we obtain the \emph{shifted} correlation function,
\begin{equation}
  \tilde{C}(t,\delta t) = C(t)-C(t+\delta t) \, , \label{eq:shifted_pi_corr}
\end{equation}
where $\delta t$ is a positive offset of one or more time slices.
By virtue of the subtraction, this modified correlation function does not have a constant contribution.
For large enough source sink separations, we may assume that it is dominated by the ground state, resulting in a time dependence of the form
\begin{equation}
 \tilde{C}(t,\delta t) = A \left[ e^{-Mt}(1 - e^{-M\delta t} ) + e^{-M(T-t)}( 1 - e^{M\delta t}) \right] \, ,
\end{equation}
where $T$ is the time extent of the lattice.
As an example, the shifted charged and full neutral pion correlation functions, computed on ensemble \emph{cA2.09.48}, are shown in Figure~\ref{fig:shifted_pi_corr}.
An appropriate effective mass for a correlation function of this type can be obtained by numerically solving, at each source sink separation, the ratio
\begin{equation}
  \frac{\tilde{C}(t+1,\delta t)}{\tilde{C}(t,\delta t)} = \frac{\left[ e^{-M(t+1)}(1 - e^{-M\delta t} ) + e^{-M(T-t-1)}( 1 - e^{M\delta t}) \right]}{\left[ e^{-Mt}(1 - e^{-M\delta t} ) + e^{-M(T-t)}( 1 - e^{M\delta t}) \right]}  \, , \label{eq:shifted_effmass}
\end{equation}
for the unknown $M$.
The effective mass for the full neutral pion as well as the correlated difference of the effective masses of the charged and full neutral pions for ensemble \emph{cA2.09.48} are shown in the left and right panels of Fig.~\ref{fig:shifted_pi_effmasses}, respectively.
Within the large statistical errors, it seems that the mass splitting is indeed either very slightly positive or compatible with zero.

In order to ascertain whether the method presented above is reliable, we have confirmed that on our old $N_f=2+1+1$ ensembles which do not employ the clover term, the masses and (significant) mass splittings are fully consistent with the results obtained in Ref.~\cite{Herdoiza:2013sla} using the traditional analysis technique.

\subsection{Fit Range Dependence and Reliable Central Values}\label{app:fitrange}

\begin{figure}
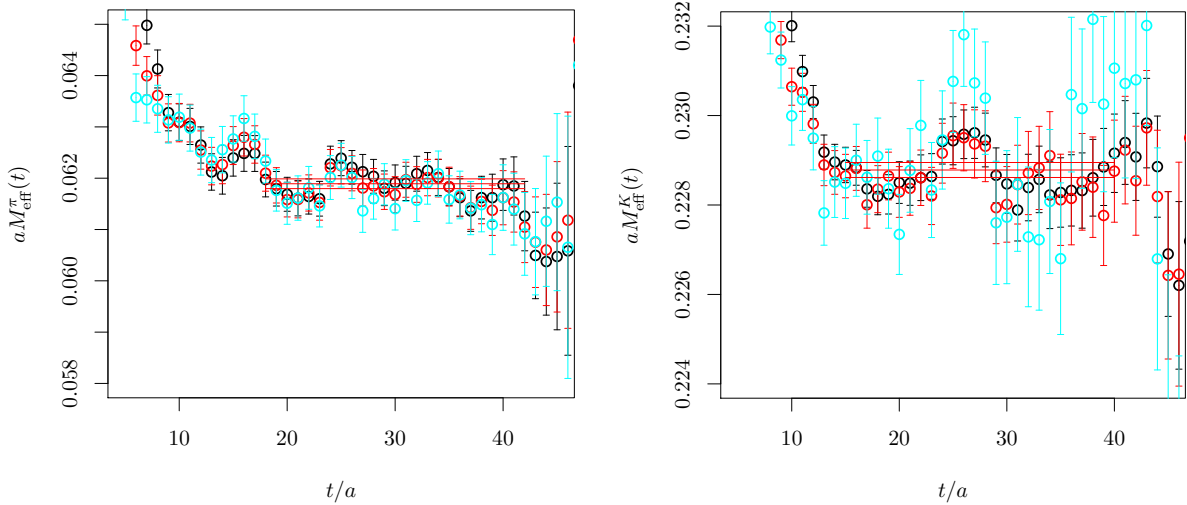

 \centering
 \includegraphics[width=0.46\textwidth]{{{m_pi.effectivemass}}} \hspace{0.3cm}
 \includegraphics[width=0.46\textwidth]{{{m_K.effectivemass}}}
 \caption{Effective masses from the pion (left) and kaon (right) correlation functions for local-local (black), local-fuzzed (red) and fuzzed-fuzzed (cyan) quark propagator contractions. Oscillations and correlations between time-slices can clearly be seen here, affecting the apparent locations of the onset of the respective plateaus. The three horizontal lines indicate one possible choice of fit range and the resulting effective mass and statistical errors of fitting a constant to the effective mass plateau, making use of the full inverse variance-covariance matrix.}
 \label{fig:m_pi_K.effectivemass}
\end{figure}

The choice of fit range for a correlation function is rather ambiguous as excited state contamination as well as random oscillations in the data can move the apparent onset of the plateau in effective masses by multiple time-slices.
In addition, correlations between the time-slices can cause data at several successive source sink separations to rise and fall together, delaying or expediting the onset of an apparent plateau.
This kind of correlation can be seen in the effective masses of the pion and kaon shown in Figure~\ref{fig:m_pi_K.effectivemass}, for example.
Both of these effects have been studied to a limited extent as early as in Ref.~\cite{Aoki1996354}, but even modern analyses often only take into account variations of the fit ranges by a few time-slices in either direction, concluding that the resulting effect is covered by the statistical error.
Although this is true in most cases, we see that for certain quantities the spread is the level of the statistical error and we believe it should be quoted as an additional source of uncertainty.

We observe further, especially with the twisted mass clover action, that rounding errors in the computation of heavy quark propagators can induce unwanted systematic effects in the extracted masses and amplitudes.
We observe this in the case of the $D$ and $D_s$ mesons, for which the effective masses show a deviation from the plateau for large source-sink separations which significantly exceeds the statistical error.
In addition, statistical errors for large source sink separations can bias the fit result.
When fits make use of the full inverse variance-covariance matrix, enhanced decorrelation at large source sink separations can increase the relative contribution of these data to the correlated $\chi^2$ function.

\begin{figure}
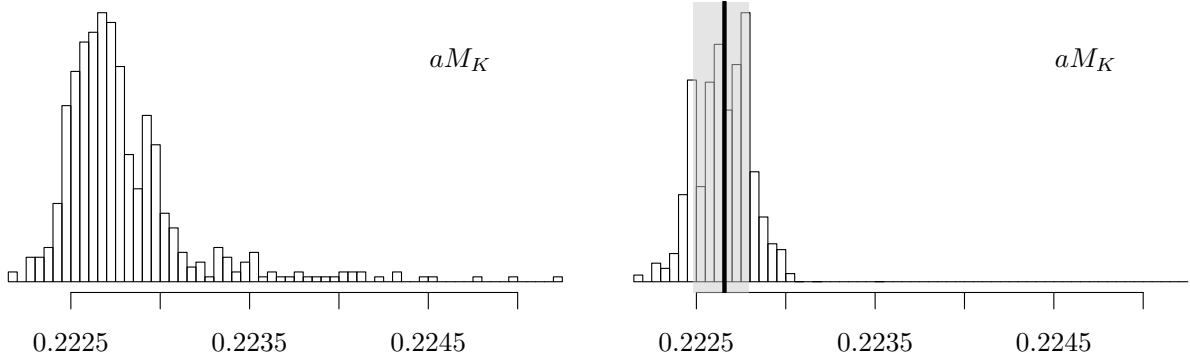

 \centering
 \includegraphics[width=0.45\textwidth,page=1]{{{fitrange.syserr.dist.m_K}}}
 \hspace{0.04\textwidth}
 \includegraphics[width=0.45\textwidth,page=2]{{{fitrange.syserr.dist.m_K}}}
 \caption{{\bf (left)} Distribution of masses extracted from constrained matrix fit to kaon correlation function at $a\mu_s=0.0245$ on the set of all ``reasonable'' fit ranges as described in the body of the text. {\bf(right)} The same distribution after weighting according to Eq.~\ref{eq:fr_weight} with the weighted median indicated by the thick vertical line and 34.27 percentiles around the median shown by the shaded area. }
 \label{fig:kaon_fitrange}
\end{figure}

In order to quantify the uncertainties entailed by the freedom to choose the fit range, we have attempted a somewhat novel analysis technique which takes into account all possible (reasonable) fit ranges, as already applied in Ref.~\cite{Helmes:2015gla}.
We make a somewhat arbitrary choice of about $0.5~\mathrm{fm}$ for the minimum length of a fit range corresponding to 6 successive time-slices ($(\Delta t)_\mathrm{min}$) and define ``reasonable'' to mean that all fits are required to converge on all fit ranges on all bootstrap samples in the analysis.
A complete listing of the minimum and maximum source-sink separations for the various quantities in this analysis is given in Table~\ref{tab:ps_params} on page~\pageref{tab:ps_params}.
In the case of the kaon on the \textit{cA2.09.48} ensemble at the physical pion mass, for example, this results in 561 fits with a distribution of fitted masses as shown in the left panel of Figure~\ref{fig:kaon_fitrange}.
Subsequently the fits are weighted according to their $p$-values and statistical errors, $\Delta$, by the weight
\begin{equation}
 \label{eq:fr_weight}
 w=\left( \frac{1}{\Delta} \left( 1-2\cdot\left|p - 0.5\right| \right) \right)^2\,,
\end{equation} 
resulting in the distribution in the right panel of Figure~\ref{fig:kaon_fitrange}.

The same approach is taken for ratios of observables, with the difference that only those fit ranges are considered which have been taken into account for both the dividend and the divisor and the weight is taken to be
\begin{equation}
 \label{eq:fr_ratio_weight}
 w=\left( \frac{1}{\Delta_{12}} \right)^2 \left( 1-2\cdot\left|p_1 - 0.5\right| \right) \left( 1-2\cdot\left|p_2 - 0.5\right| \right)\,.
\end{equation}

The median of this weighted distribution is taken as the central value and its statistical error is computed on the bootstrap samples.
The estimate of the systematic error is given by the 34.27 percentiles around the median.
As an example, the median and systematic error in the determination of $aM_K$ at $a\mu_s=0.0245$ are shown in Figure~\ref{fig:kaon_fitrange} and can be taken as an indication of what can be expected on the final results.

\subsection{Tuning the Strange and Charm Valence Quark Masses}\label{app:mass_tuning}

In Ref.~\cite{Abdel-Rehim:2013yaa}, the bare strange and charm valence quark masses were fixed using the $N_f=2$ light-strange quark mass ratio from Ref.~\cite{Aoki:2013ldr} and the strange-charm quark mass ratio given in Ref.~\cite{Davies:2009ih}, which we will call the FLAG and HPQCD quark mass ratios respectively.
Although this is valid because we are working at the physical point, the quark mass ratios in the literature have significant uncertainties and simply using their central values to set the valence masses does not allow to propagate the resulting uncertainty to observables.

In order to obtain a reliable error estimate and to allow comparison with an analysis performed using quark mass ratios, the method used in the present work consists of computing all quantities at four values each of the bare strange and charm quark masses on the ensemble with the physical light quark mass, resulting in up to 16 different combinations for observables depending on both, as listed in Table~\ref{tab:ps_params} on page~\pageref{tab:ps_params}.
At heavier than physical light quark mass, we use a total of 36 combinations of strange and charm masses in order to cover a larger range and to make more reliable interpolations.

\begin{figure}
 \includegraphics[width=0.49\linewidth]{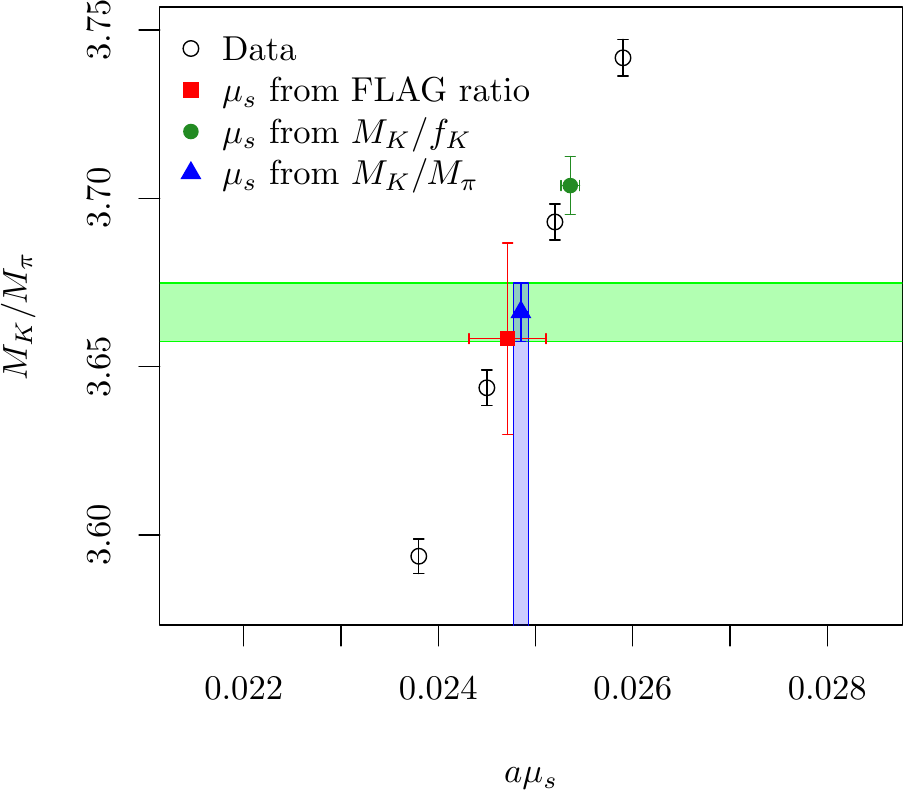}
 \includegraphics[width=0.49\linewidth]{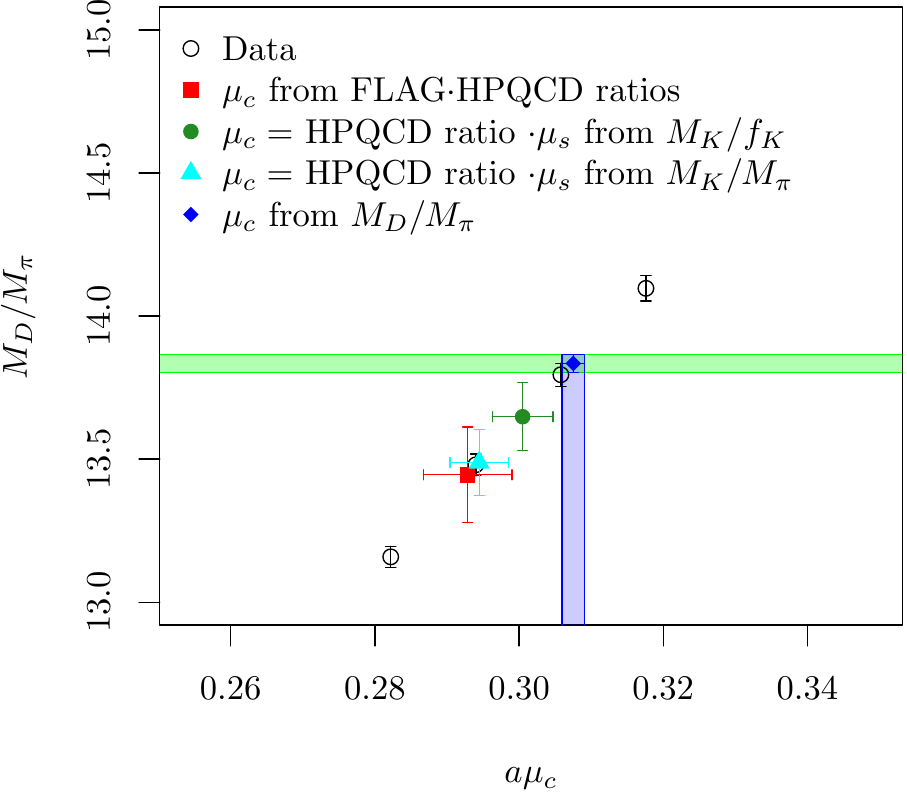}
 \caption{Linear interpolation in the strange and charm quark masses and matching with the phenomenological values of the ratios $M_K/M_\pi$ (blue triangle in the left-hand panel) and $M_D/M_\pi$ (blue diamond in the right-hand panel). The other points are given for reference and of these, the strange quark mass from $M_K/f_K$ is determined using the ``continuum definition'' of $af_K$.}
  \label{fig:sc_interp}
\end{figure}

We then perform a linear fit in the valence quark mass dependence and can subsequently interpolate to match the bare quark masses determined from the quark mass ratios above with the error properly propagated.
Alternatively, we can fix the strange and charm quark masses by interpolating to match the phenomenological values of some ratios of mesonic quantities.
Here, we do this for $M_K/M_\pi$ and $M_D/M_\pi$ as shown in Figure~\ref{fig:sc_interp}, resulting in much smaller errors on the strange and charm quark masses in the rest of the analysis than by using quark mass ratios and propagating their uncertainties or by matching $M_K/f_K$.
The bare quark masses determined from the different methods are given in the Table~\ref{tab:quarkmasses}, where the first error is statistical the second error stems from the analysis described above in Appendix~\ref{app:fitrange}.
All the final results are quoted for $a\mu_s$ and $a\mu_c$ as determined from the matching to the phenomenological values of $M_K/M_\pi$ and $M_D/M_\pi$ with the other values given for comparison only.

\begin{table}
  \centering
\begin{center}
  \vspace{0.2cm}
  \noindent{\small
  \begin{tabular*}{0.8\textwidth}{@{\extracolsep{\fill}}lll}
    \hline \hline \\[-2.2ex]
                              & $a\mu_s$                              & $a\mu_c$ \\[0.6ex]
    \hline \\[-2.0ex]
    FLAG/HPQCD                & $0.0247(4)$                           & $0.293(6)^\star$ \\[0.6ex]
    $M_K/f_K^\mathrm{(CD)}$   & $0.02536(10)(^{+05}_{-05})$           & $0.3005(42)(^{+06}_{-06})^\star$ \\[0.6ex]
    $M_K/f_K^\mathrm{(LD)}$   & $0.02480(10)(^{+04}_{-04})$           & $0.2938(41)(^{+04}_{-05})^\star$ \\[0.6ex]
    $M_K/M_\pi$               & $\mathbf{0.02485(7)(^{+4}_{-3})}$     & $0.2940(40)(^{+04}_{-04})^\star$ \\[0.6ex]
    $M_D/f_D^\mathrm{(CD)}$   & --                                    & $0.3629(66)(^{+70}_{-96})$ \\[0.6ex]
    $M_D/f_D^\mathrm{(LD)}$   & --                                    & $0.2902(26)(^{+09}_{-17})$ \\[0.6ex]
    $M_D/M_\pi$               & --                                    & $\mathbf{0.3075(15)(^{+14}_{-14})}$ \\[0.6ex]
    \hline \hline
  \end{tabular*} }
  \vspace{0.2cm}
\end{center}
  \caption{Bare quark masses resulting from matching the the quantity in the leftmost column. The labels (LD) and (CD) correspond to $f_K$ ($f_D$) extracted according to the two definitions given in Eq.~\eqref{eq:fps}. The starred $a\mu_c$ are derived from the corresponding $a\mu_s$ and the HPQCD charm to strange ratio. The bold values are the strange and charm quark masses used for the final results of the analysis.}
  \label{tab:quarkmasses}
\end{table}

It is interesting to note that for the strange quark mass the use of the lattice definition of the decay constant in the ratio $M_K/f_K$ results in good agreement with the value of $a\mu_s$ as given by the $N_f=2$ FLAG strange to light quark mass ratio and the one determined from $M_K/M_\pi$.
In the charm sector, using the lattice definition in $M_D/f_D$ results in a charm quark mass which agrees with the values determined via the HPQCD charm to strange quark mass ratio and the three strange quark masses discussed above.
The statistical and systematic errors on $a\mu_c$ derived from the lattice definition of $M_D/f_D$ are quite small because in this definition the charm quark mass dependence of $af_D$ is suppressed, giving $M_D/f_D$ a substantial slope.
The large value of $a\mu_c$ and the associated uncertainties derived from the continuum definition of $M_D/f_D$ just reflect how flat the behaviour of this ratio becomes as a function of $a\mu_c$ in this case, which can be seen as an indication for discretisation errors.
The charm quark mass determined from $M_D/M_\pi$ has a statistical uncertainty lower by a factor of two or three compared to the other estimates but disagrees with their values.
In addition to the possibly sizeable lattice artefacts in $M_D$, finite size corrections on $M_\pi$ are likely to be at the few percent level which means that without the necessary corrections, the current uncertainties are likely to be strongly underestimated.

In principle, at the cost of losing predictivity for $f_K$ and $f_D$, estimates for the physical strange and charm quark masses could be derived from weighted averages of some or all of the lattice determinations given in Table~\ref{tab:quarkmasses}.
The spread of the different values could then be taken as a first estimate of systematic uncertainties due to discretisation and finite volume artefacts.
In addition, similar to what was done in section~\ref{subsec:lattice_spacing} for the lattice spacing, determinations from the baryon sector could be used to increase confidence in the quark mass estimates.

\subsection{Interpolations} \label{app:interpolations}

\begin{figure}
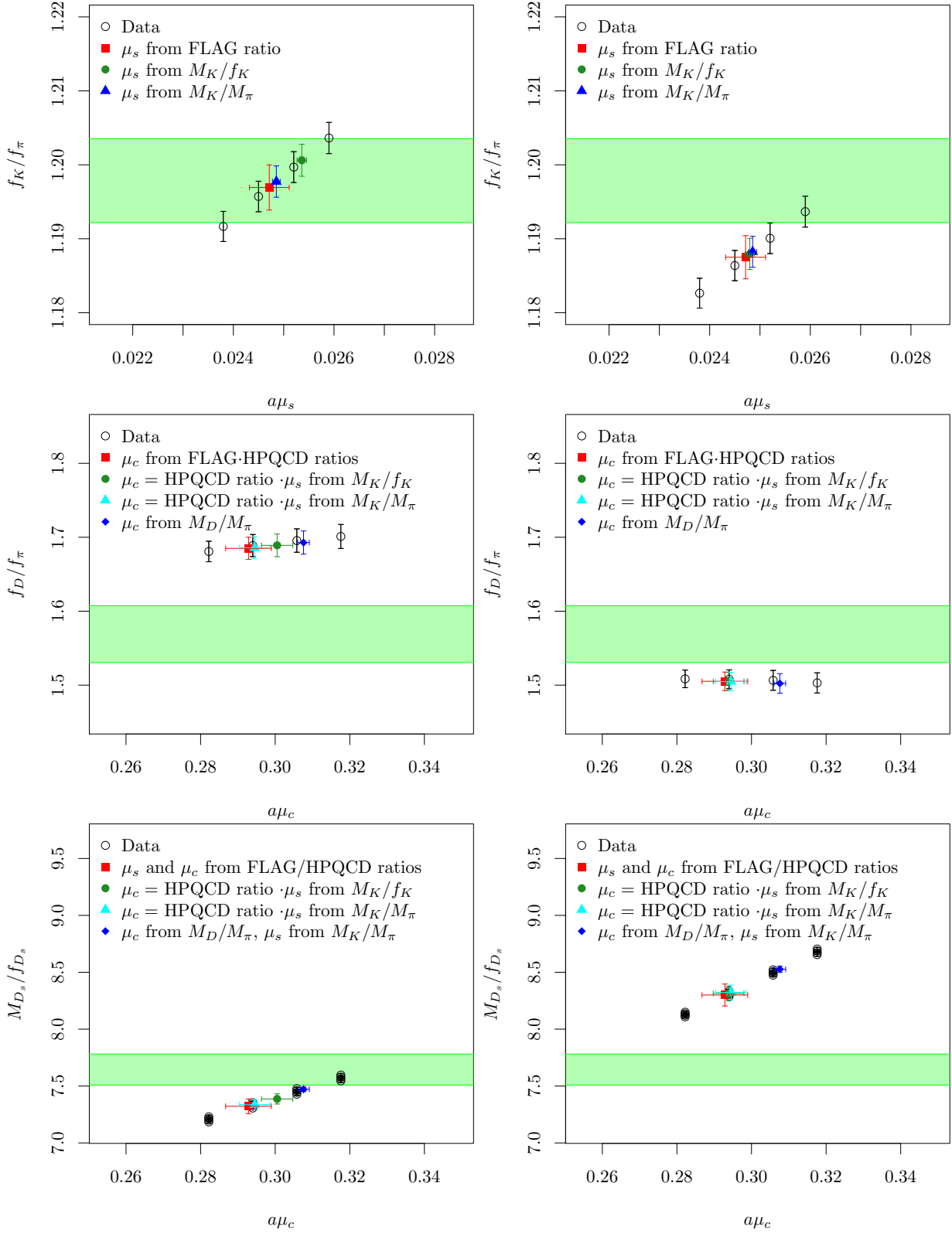

  \centering
  \includegraphics[width=0.49\linewidth]{{{f_K_ov_f_pi_continuum}}}
  \includegraphics[width=0.49\linewidth]{{{f_K_ov_f_pi_lattice}}}
  \includegraphics[width=0.49\linewidth]{{{f_D_ov_f_pi_continuum}}}
  \includegraphics[width=0.49\linewidth]{{{f_D_ov_f_pi_lattice}}}
  \includegraphics[width=0.49\linewidth]{{{m_Ds_ov_f_Ds_continuum}}}
  \includegraphics[width=0.49\linewidth]{{{m_Ds_ov_f_Ds_lattice}}}
\caption{Representative choice of interpolations of various quantities involving decay constants using the continuum definition {\bf (left)} and the lattice definition {\bf (right)} with the phenomenological value indicated by the green band.}
\label{fig:rep_interp}
\end{figure}

After reliable central values, statistical and systematic errors due to fit range arbitrariness have been determined for the observables as described above, we perform independent linear interpolations in all quantities under study towards the values of the strange and charm quark masses listed in Table~\ref{tab:quarkmasses}.
In principle other approaches could be used for the interpolations, such as using the squares of the quantities or forms inspired by chiral perturbation theory, but for reasons of simplicity and consistency all the data are interpolated linearly.
This seems to be very well justified by the small mass range in the interpolations and the shape of the quark mass dependences.
Here, only the statistical error is used as a weight for the linear fits because of the difficulties involved in defining a sum of squared residuals with asymmetric weights. 
The statistical uncertainties in the values of the quark masses are propagated by Taylor expansion, contributing to the total statistical error of the interpolation results in quadrature.
Illustrations of a representative set of these interpolations are shown in Figure~\ref{fig:rep_interp} with the continuum definition of the decay constant shown in the left panels and the lattice counterpart in the right panels.

\begin{figure}
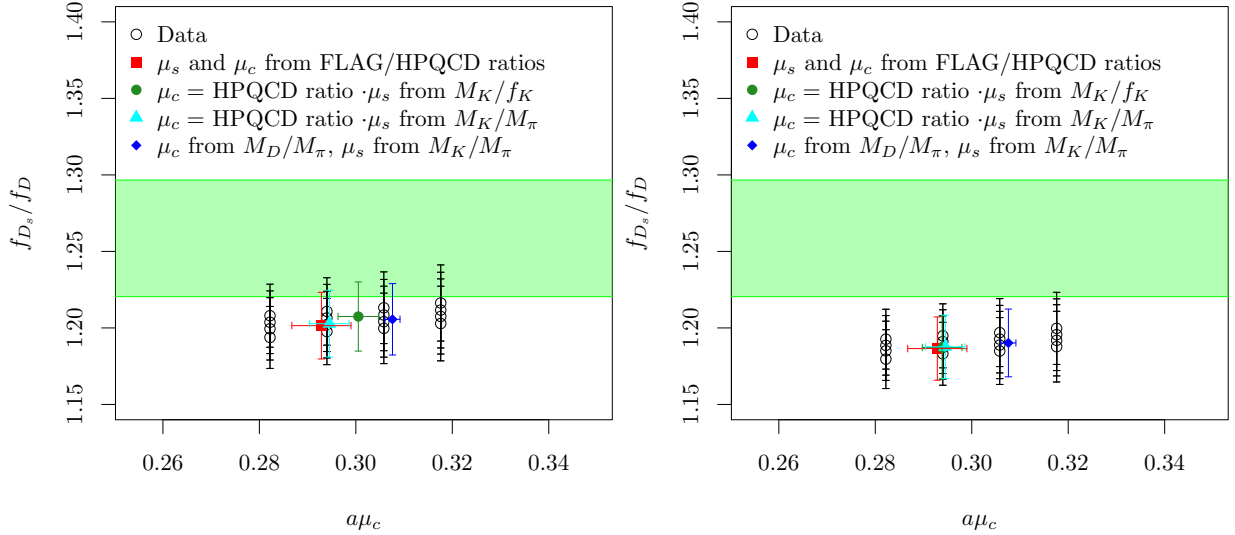

  \includegraphics[width=0.49\linewidth]{{{f_Ds_ov_f_D_continuum}}}
  \includegraphics[width=0.49\linewidth]{{{f_Ds_ov_f_D_lattice}}}
 \caption{Quark mass interpolation of the ratio $f_{D_s}/f_D$ on the physical pion mass ensemble \textit{cA2.09.48} with the phenomenological value indicated by the green band. (left) Continuum definition of the decay constants. (right) Lattice definition of the decay constants. }
 \label{fig:interp_f_Ds_ov_f_D}
\end{figure}

A number of features seen in Figure~\ref{fig:rep_interp} deserve discussion.
The first one we would like to mention is the fact that for many of the quantities that were analysed, the quark mass dependence is so weak that within errors, no appreciable slope is observed over the whole range of strange and charm quark masses.
Consequently the error is also largely independent of which set of strange and charm quark masses is used, as exemplified by the quark mass interpolation of the ratio $f_{D_s}/f_D$ in Figure~\ref{fig:interp_f_Ds_ov_f_D}.
This is of course not unexpected for decay constants and it shows that for many quantities, slight mis-tuning of the valence strange and charm quark masses does not lead to significant biases when working at the physical light quark mass.
On the other hand, for a quantity like $M_{D_s}/f_{D_s}$, which has a noticeable slope in both the strange and charm mass, it has a strong effect  on the central value as well as the errors which interpolation point is chosen.
Errors differ by up to a factor of 4 as shown in the bottom-most panels of Figure~\ref{fig:rep_interp}.
The next notable feature concerns the (unsurprisingly) rather large effect of the definition of the decay constant on ratios involving $f_D$ and $f_{D_s}$, except when both are involved simultaneously.
Clearly, the two definitions agree in the continuum limit and these differences show that discretisation effects could be at the level of 15\% to 20\% for quantities involving charm quarks.
Finally, quantities involving $M_\pi$ and $f_\pi$ are expected to be subject to finite volume corrections at the few-percent level which will be accounted for in a future study with finer lattice spacings and larger volumes.

\begin{figure}
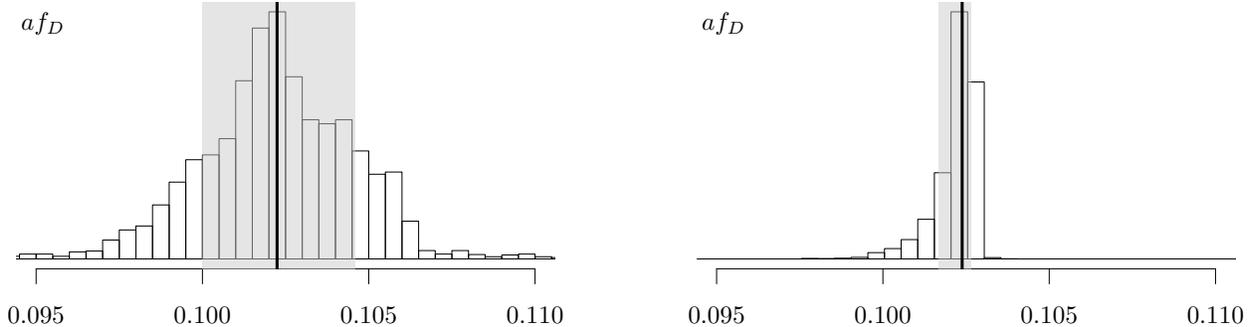

 \centering
 \includegraphics[width=0.45\textwidth]{{{f_D.interp.hist.unweighted}}} \hfill
 \includegraphics[width=0.45\textwidth]{{{f_D.interp.hist.weighted}}}
 \caption{Propagation of the systematic error due to fit range ambiguity to the interpolation of $af_D$. {\bf (left):} Distribution resulting from uniform sampling of data from different fit ranges. {\bf (right):} Distribution resulting from sampling which accounts for the weights of different data points as relative sampling probabilities. The median and the 34.27 percentiles around the median, our estimate of the systematic error, are indicated by the thick vertical line and the grey rectangle.}
 \label{fig:f_D_interpolation_systematic_error}
\end{figure}

To propagate the systematic error to the quark mass estimates and the results of interpolations, we generate 5000 random samples of the data points involved in a given interpolation by randomly drawing from the various fit ranges for each combination of quark masses.
In order to obtain a reliable estimate of the resulting error, rather than sampling uniformly, we use the weights from Eqs.~\ref{eq:fr_weight} and \ref{eq:fr_ratio_weight} as relative sampling probabilities, such that data with large weights occurs more frequently in the set.
The effect of this choice can be rather profound as we show in Figure~\ref{fig:f_D_interpolation_systematic_error} for $af_D$, where the left panel corresponds to the distribution when the data from different fit ranges is sampled uniformly and the right panel shows the distribution when the weights are taken into account.
It must be noted that the weighted distribution corresponds to what is observed for $af_D$ at the four charm quark masses, justifying the approach.

\begin{figure}
 \centering
 \includegraphics[width=0.75\textwidth]{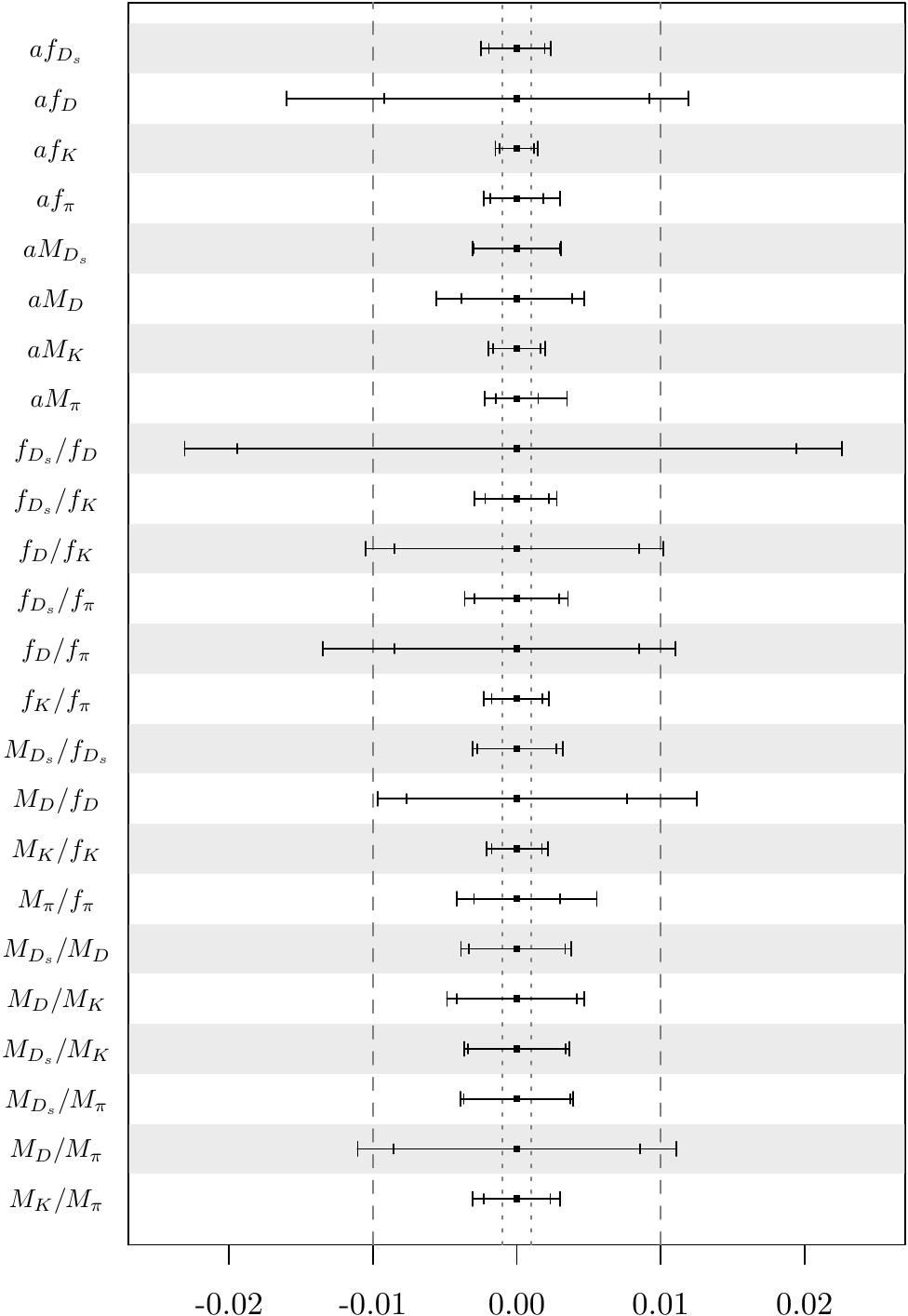}
 \caption{Error budget for various mesonic observables on ensemble \textit{cA2.09.48} relative to their central values. Observables involving strange and charm quarks have been interpolated to physical strange and charm quark masses as described in Appendix~\protect{\ref{app:interpolations}}. The inner error bar is statistical and includes contributions from the fitting procedure as well as the error propagated from the uncertainty in the quark mass estimates. The outer error bar indicates the systematic error due to the ambiguity in the choice of fit range. The dotted and dashed lines show the per-mille and percent error boundaries respectively. The errors are shown cumulatively and would add in quadrature if combined.}
 \label{fig:interpolations_error_summary}
\end{figure}

A summary of the statistical and systematic errors is given in Figure~\ref{fig:interpolations_error_summary} for the 24 quantities from this analysis, normalised by their respective central values and including those that are technically not independent.
It is clear that for most quantities the choice of fit range has a very limited effect on the total uncertainty and past analyses were probably well-justified in using only one or a few fit ranges, despite what appears to be a significant ambiguity involved in the choice of fit range.
For the pion mass and decay constant, however, the systematic error is of the same order as the statistical error and must be taken into account.

Quantities involving the D meson show significant spread which might increase even further as the volume is enlarged and more fit ranges become available.
On the one hand this is due to the light pion mass, which limits the signal to noise ratio for large source-sink separations in observables involving light quarks.
On the other hand, we observe deviations in the effective mass from a plateau which exceed the statistical error and these are likely due to increased round-off errors in the computation of heavy quark propagators at coarse lattice spacings and with the clover term present.
Since these round-off errors limit the maximum source-sink separations that can be taken into account, we would like to perform future studies with more robust solvers, possibly with a number of iterations in quadruple precision, in order to significantly extend the source-sink separation for which a reasonable plateau can be observed as suggested in Ref.~\cite{Juttner:2005ks}.
Finally we would like to note that for many quantities, per-mille level uncertainties are likely not attainable by working just with simulations at the physical pion mass, because the signal to noise ratio deteriorates as the light quark mass is reduced to its physical value.
As a result, a procedure involving simulations at heavier than physical light quark masses in combination with (heavy meson) Wilson chiral perturbation theory, stabilised by measurements at the physical point, may be necessary.
This finding is in line with what was already observed in Refs.~\cite{PhysRevD.91.054507} and~\cite{PhysRevD.90.074509}.

\begin{table}
 \small
 \begin{tabular*}{1.\textwidth}{@{\extracolsep{\fill}}llllllll}
    \hline \hline
  $L/a$   & \multicolumn{7}{c}{bare valence quark masses} \\
  \hline \\[-2.0ex]
  \multirow{3}{*}{24, 32} & $a\mu_l$ &          & $0.003$   & $0.006$  &          &          & \\
                          & $a\mu_s$ & $0.0224$ & $0.0231$ & $0.0238$ & $0.0245$ & $0.0252$ & $0.0259$ \\
                          & $a\mu_c$ & $0.2586$ & $0.2704$ & $0.2822$ & $0.294$  & $0.3058$ & $0.3176$ \\
  \hline
  \multirow{3}{*}{48}     & $a\mu_l$ & $0.0009$ &          &          &          &          & \\
                          & $a\mu_s$ &          & $0.0231$ & $0.0238$ & $0.0245$ & $0.0252$ & \\
                          & $a\mu_c$ &          & $0.2704$ & $0.2822$ & $0.294$  & $0.3058$ & \\
  \hline \hline \\
 \end{tabular*}
 \begin{tabular*}{1.\textwidth}{@{\extracolsep{\fill}}lccccccc}
  \hline \hline \\[-2.0ex]
  $L/a$                   & \multicolumn{6}{c}{fit range minimum and maximum time-slices} \\[0.6ex]
  \hline \\[-2.0ex]
                          & $\pi^\pm$ & ${\pi^{(0,c)}}^\star$ & $\pi^0$   & $K$      & $D$       & $D_s$ & $(\Delta t)_\mathrm{min}$ \\
  24                      & $[9,23]$  & $[9,23]$              & $[7,23]$  & $[9,23]$ & $[12,23]$ & $[18,23]$ & 6 \\
  32                      & $[9,28]$  & $[9,28]$              & $[7,31]$  & $[9,28]$ & $[13,27]$ & $[15,27]$ & 6 \\
  48                      & $[9,47]$  & $[9,47]$              & $[7,47]$  & $[9,47]$ & $[11,35]$ & $[11,35]$ & 6 \\
  \hline \hline
 \end{tabular*}
 \caption{Bare valence quark mass parameters and fit range restrictions for the computation of pseudo-scalar meson correlators used in this analysis. $\star$ : $(0,c)$ refers to the connected part of the neutral pion }
 \label{tab:ps_params}
\end{table}

\end{document}